\documentclass[twocolumn]{aastex63}
\usepackage[utf8]{inputenc}
\usepackage{CJKutf8}
\usepackage{url}
\usepackage{amsmath}
\usepackage{amssymb}
\usepackage{bm}
\usepackage[toc,page]{appendix}
\usepackage{listings}

\newcommand\numberthis{\addtocounter{equation}{1}\tag{\theequation}}

\begin{document}

\title{Improving Power Spectrum Estimation using Multitapering:\\
Efficient asteroseismic analyses for understanding stars, the Milky Way, and beyond}

\author[0000-0002-7626-506X]{Aarya A. Patil}
\correspondingauthor{Aarya A. Patil}
\email{patil@mpia.de}
\altaffiliation{LSST-DA Catalyst Fellow}
\affiliation{Max-Planck-Institut für Astronomie, Königstuhl 17, D-69117 Heidelberg, Germany}
\affiliation{David A. Dunlap Department of Astronomy \& Astrophysics, University of Toronto, 50 St George Street, Toronto ON M5S 3H4, Canada}
\affiliation{Dunlap Institute for Astronomy \& Astrophysics, University of Toronto, 50 St George Street, Toronto, ON M5S 3H4, Canada}

\author[0000-0003-3734-8177]{Gwendolyn M. Eadie}
\affiliation{David A. Dunlap Department of Astronomy \& Astrophysics, University of Toronto, 50 St George Street, Toronto ON M5S 3H4, Canada}
\affiliation{Department of Statistical Sciences, University of Toronto, 9th Floor, Ontario Power Building, 700 University Ave, Toronto, ON M5G 1Z5, Canada}

\author[0000-0003-2573-9832]{Joshua S. Speagle (\begin{CJK*}{UTF8}{gbsn}沈佳士\ignorespacesafterend\end{CJK*})}
\affiliation{Department of Statistical Sciences, University of Toronto, 9th Floor, Ontario Power Building, 700 University Ave, Toronto, ON M5G 1Z5, Canada}
\affiliation{David A. Dunlap Department of Astronomy \& Astrophysics, University of Toronto, 50 St George Street, Toronto ON M5S 3H4, Canada}
\affiliation{Dunlap Institute for Astronomy \& Astrophysics, University of Toronto, 50 St George Street, Toronto, ON M5S 3H4, Canada}
\affiliation{Data Sciences Institute, University of Toronto, 17th Floor, Ontario Power Building, 700 University Ave, Toronto, ON M5G 1Z5, Canada}

\author{David J. Thomson}
\altaffiliation{FRSC \& Emeritus Professor}
\affiliation{Department of Mathematics \& Statistics, Queen's University, Kingston, ON K7L 3N6, Canada}

\begin{abstract}
Asteroseismic time-series data have imprints of stellar oscillation modes, whose detection and characterization through time-series analysis allows us to probe stellar interior physics. Such analyses usually occur in the Fourier domain by computing the Lomb-Scargle (LS) periodogram, an estimator of the \textit{power spectrum} underlying unevenly-sampled time-series data. However, the LS periodogram suffers from the statistical problems of (1) inconsistency (or noise) and (2) bias due to high spectral leakage. Here, we develop a multitaper power spectrum estimator using the Non-Uniform Fast Fourier Transform (\texttt{mtNUFFT}) to tackle the inconsistency and bias problems of the LS periodogram. Using a simulated light curve, we show that the \texttt{mtNUFFT} power spectrum estimate of solar-like oscillations has lower variance and bias than the LS estimate. We also apply our method to the Kepler-91 red giant, and combine it with \texttt{PBjam} peakbagging to obtain mode parameters and a derived age estimate of $3.97 \pm 0.52$ Gyr. \texttt{PBjam} allows the improvement of age precision relative to the $4.27 \pm 0.75$ Gyr APOKASC-2 (uncorrected) estimate, whereas partnering \texttt{mtNUFFT} with \texttt{PBjam} speeds up peakbagging thrice as much as LS. This increase in efficiency has promising implications for Galactic archaeology, in addition to stellar structure and evolution studies. Our new method generally applies to time-domain astronomy and is implemented in the public Python package \texttt{tapify}, available at \url{https://github.com/aaryapatil/tapify}.
\end{abstract} 

\section{Introduction} \label{sec:intro}
Modern advances in the theory of stellar structure and evolution are driven by high-precision photometric time-series observed using space-based telescopes such as MOST \citep{walker_2003}, CoRoT \citep{baglin_2009, auvergne_2009}, Kepler \citep{borucki_2010, koch_2010} (and K2), BRITE \citep{weiss_2014}, TESS \citep{ricker_2015}, and the upcoming PLATO mission \citep{rauer_2014} \citep[e.g.,][]{buzasi_2000, michel_2008, miglio_2009, aerts_2010_book, deridder_2009, degroote_2010, chaplin_2011, li_2020}.

Analyses of these observations in the Fourier domain exhibit the frequencies at which stars oscillate. By studying these frequencies, asteroseismology provides a unique pathway to investigate the deep interior of stars and the physical mechanisms that drive oscillations.

To obtain Fourier domain representations of stellar oscillations, one estimates the power spectrum from the light curve. In the power spectrum, features at different frequencies are associated with different physical phenomena, and these features in turn depend on the type of pulsating star \citep[refer to the pulsation HR diagram in][chapter 2]{aerts_2010_book}. In the case of stars that have solar-like oscillations, we can observe several features in the power spectrum in addition to the peaks associated with oscillation modes. These power spectrum features are described as follows \citep{garcia_2019}:  
\begin{enumerate}
    \item rotational modulation peaks and harmonics,
    \item continuum resulting from granulation,
    \item pressure (p) mode envelope of resonant oscillations, \label{item:p-mode}
    \item and a photon noise level.
\end{enumerate} 
Together, these features provide the most stringent constraints on stellar structure models. In addition, space-based photometry enables high-precision measurements of the orbital frequencies and harmonics of transiting exoplanets, allowing us to better model planetary systems.

Solar-like oscillations are expected in stars with convective envelopes. We thus observe them in low-mass main sequence ($M \lesssim 1.5 M_\odot$), subgiant branch, and G-K red giant stars \citep{hekker_2011, white_2011}, which form the most abundant type of oscillators. A set of acoustic p-modes or standing sound waves probe the turbulent outer layers as well as the deep interior of these oscillators (refer to point \ref{item:p-mode} above). In theory, these modes are damped, stochastically excited harmonic oscillations, represented by a sequence of Lorentzian profiles with nearly even frequency spacing \citep{aerts_2010_book}. We can characterize these modes in power spectra to estimate stellar masses and radii using either a model-independent or model-dependent approach. The model-independent approach uses simple scaling relations with the Sun \citep{kjeldsen_1995} and is efficient as compared to detailed stellar modeling. However, its accuracy and precision is limited by that of $\Delta \nu$ and $\nu_\mathrm{max}$ estimates and the approximations underlying the scaling relations. The stellar model-dependent approach \citep[e.g.,][]{miglio_2005, quirion_2010, silva-aquirre_2017} provides more accurate and precise estimates, with the frequency estimates being the major source of uncertainty.

In this paper, we target the efficient extraction of $\Delta \nu$ and $\nu_\mathrm{max}$ as well as individual p-mode parameters as a way to provide stringent constraints on stellar masses, radii, and therefore ages \citep{bellinger_2019}. We present a new frequency analysis method, the multitaper NUFFT (\texttt{mtNUFFT}) periodogram, that mitigates the statistical issues of the standard LS periodogram, resulting in better power spectrum estimates (detailed in Section \ref{subsec:stats}). This new periodogram is an extension of the \texttt{mtLS} periodogram developed in \cite{springford_2020}. Our focus is mainly on mode parameter estimation for red giants, as these stars help characterize ensembles of stellar populations out to large distances, thereby enabling Galactic archaeological studies.

In addition to inference of stellar properties, space-based light curve data embed information of exoplanets orbiting stars \citep{auvergne_2009, borucki_2010, ricker_2015}. Accurate estimation of the fundamental properties of stars such as mass, radius, and age allows better characterization of the exoplanets they host. Therefore, improvements in asteroseismology help resolve outstanding questions on the formation and evolution of planetary systems.

To estimate power spectra, \texttt{mtNUFFT} computes and stores complex Fourier coefficients. The phase information in these coefficients allows us to perform a hypothesis test called Thomson's Harmonic \textit{F-test} \citep{thomson_1982}, which determines whether the power at a given frequency is predominantly due to a single strictly periodic, sinusoidal-shaped signal. The LS periodogram and its multitaper version \citep[\texttt{mtLS};][]{springford_2020}, on the other hand, do not readily offer such phase information. Thus, \texttt{mtNUFFT} enables harmonic analysis via the F-test while the LS and \texttt{mtLS} do not. In our accompanying paper, Patil et al. (2024b), we show that the F-test allows us to detect pure sinusoids and can therefore be used to investigate two types of asteroseismic modes: gravity (g) modes and undamped modes with quasi-infinite lifetimes. This paper thus demonstrates that the \texttt{mtNUFFT} along with the F-test has potential applications to other types of pulsating stars exhibiting either g or undamped modes. Note that the F-test also helps understand the transient nature of damped p-modes. We show this in Patil et al. (2024b) to highlight that the F-test can provide major improvements to estimating p-mode frequencies.

\subsection{Statistical Background}\label{subsec:stats}

In order to obtain high-precision frequency estimates of p-modes using light curve data, we need a statistically reliable estimator of the power spectrum. Many power spectrum estimators have been developed for data sampled regularly in time and their statistical properties are well established in the literature. The oldest of these, the \textit{classical periodogram} \citep{schuster_1898}, is commonly used in science and engineering but is inconsistent and biased. The inconsistency comes from non-zero variance (or noise) in the estimate. On the other hand, a biased estimator is one whose expected value is not the true underlying value, i.e., its estimates differ from the truth on average. The bias in the periodogram comes from high spectral leakage, which is power belonging to frequency $f$ that manifests at a far-away frequency $f + \Delta f$. 

While there exists no unbiased estimator of the power spectrum underlying a discrete time-series sampled over a finite observation interval, estimators that taper the data significantly reduce and control bias \citep{brillinger_1981}. A taper or a tapered window (sometimes simply referred to as window) is a weight function that is applied to data in the time-domain to reduce bias introduced by the finite observation interval \citep{harris_1978}. Mathematically, a windowed time-series is given by
\begin{equation}
    v(t) \cdot x(t)
\end{equation} where $x(t)$ is the time-series and $v(t)$ is the window, both evaluated at time $t$. An un-tapered time-series is defined by the so-called rectangular window
\begin{equation}
    v_R(t) = 
    \begin{cases}
      1 & t_0 \le t \le t_N\\
      0 & \text{otherwise}\\
    \end{cases}       
\end{equation} which exhibits a discontinuity at the observation boundaries $t_0$ and $t_N$, whereas windows such as the cosine-tapered (or Tukey) window smoothly set the time-series data to zero near its boundary. Essentially, a tapered window smooths the sharp discontinuity at the boundaries of the observation interval to reduce bias.

While tapering reduces bias, its downside is that it increases variance (or noise). Additionally, tapers generally down weight data points near the edges of a time series, which leads to some information loss. Instead of using just one taper, \cite{thomson_1982} use multiple orthogonal tapers called Discrete Prolate Spheroidal Sequences \citep[DPSS;][]{slepian_1978} to obtain an averaged estimate from a number of single-tapered estimates (refer to Section~\ref{subsubsec:mt_general_solution} for more details). This method treats both the bias and inconsistency problems, minimizes loss of information, and outperforms un-tapered and single-tapered power spectrum estimates (with or without smoothing) \citep[refer to Section~\ref{subsubsec:mt_general_solution},][]{park_1987, bronez_1992, riedel_1994, mccoy_1998, stoica_99, prieto_2007, thomson_2014, lees_1995}. It is very popular in different fields of science and engineering; particularly interesting applications are those in Earth, climate, and solar science since they have many similarities with asteroseismology \citep[for e.g.][]{park_1987, thomson_1996, thomson_2015a, thomson_2015b, chave_2019, chave_2020, mann_2021}. There are also some astronomical studies that use multitapering such as \cite{scargle_1997, feigelson_2012, vaughan_2013, romano_2017, scargle_2021, dodson-robinson_2022b}.

Time-series data in astronomy are often dependent on observational factors resulting in irregular sampling. This is true for modern space-based asteroseismic data, e.g., Kepler observations \citep{borucki_2010, koch_2010} are over Q0-Q16 quarters, each of ${\approx}3$ months duration, with data downlinks that result in gaps as well as slight uneven sampling due to conversion of evenly-sampled timestamps to Barycentric Julian Date. Here uneven sampling refers to non-uniform time intervals between observations. While one can interpolate such unevenly-sampled time-series data to a mesh of regular times \citep[e.g.][]{garcia_2014} and use estimators based on the assumption of even sampling, \cite{lepage_2009} and \cite{springford_2020} demonstrate that interpolation leads to spectral leakage by introducing power from the interpolation method itself and thus has undesirable effects on power spectrum estimates. Instead, the Lomb-Scargle (LS) periodogram \citep{lomb_1976, scargle_1982} is widely regarded as a standard solution to the power spectrum estimation problem for irregular sampling and is particularly prevalent in astronomy. However, it suffers from the same statistical issues as the classical periodogram and its spectral leakage worsens with increased irregularity of the time samples \citep{vanderplas_2018}. 

We thus develop the \texttt{mtNUFFT} periodogram\footnote{we use the term ``periodogram'' colloquially here; it should not to be confused with that of \citep{schuster_1898}} that extends the Thomson multitaper power spectrum estimate to irregular sampling and improves upon the noise and spectral leakage properties of the LS periodogram. This new periodogram extends the functionality of the \texttt{mtLS} periodogram, and is particularly favourable for detecting oscillations in space-based light curves with quasi-regular time sampling.

\begin{figure*}[t]
    \centering
    \includegraphics[width=\linewidth]{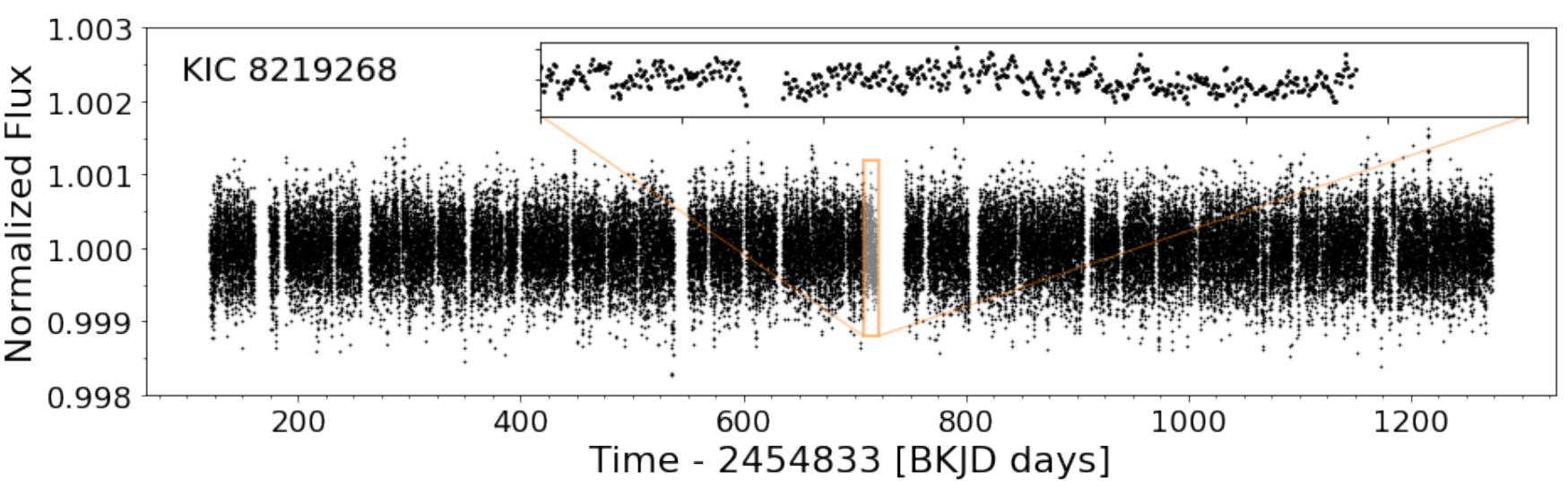}
    \caption{Photometric time series of Kepler red giant star-91. The inset shows a zoom-in plot of the time-series from 708 to 722 days, highlighting that the time sampling is uneven and that long gaps are present.}
    \label{fig:kepler_time}
\end{figure*}

\subsection{Overview}

The outline of the paper is as follows. Section \ref{sec:multitaper_math} motivates the use of multitaper power spectrum estimation in asteroseismology given its statistical background, and introduces our new method, the \texttt{mtNUFFT} periodogram. This section presents pedagogy for readers new to time-series analysis. We thus direct the experienced reader to Section \ref{subsec:multitaper} that presents our frequency analysis method and its novelty compared to the state-of-the-art. We also recommend supplementing this section with \cite{percival_1993}, an excellent educational resource on multitaper power spectrum analysis. 

To demonstrate the advantageous statistical properties of our method, we apply it to an example Kepler time-series that has p-mode oscillations: the Kepler-91 red giant \citep[][]{batalha_2013}. We also simulate a light curve of a solar-like oscillator to show that \texttt{mtNUFFT} allows better estimation of p-mode power spectra than LS. Section \ref{sec:age} applies \texttt{mtNUFFT} to our Kepler-91 case study example and combines it with peakbagging to estimate mode parameters as well as stellar mass, radius, and age. We compare these estimates with those obtained by peakbagging the LS periodogram of Kepler-91, and those from the APOKASC-2 catalog \citep{pinsonneault_2018}. We find that peakbagging with the mtNUFFT periodogram is more efficient than with LS.

Appendix \ref{sec:tapify} discusses \texttt{tapify} \citep{tapify}, a Python package we developed for multitaper power spectrum analysis, and provides a workable example. Appendix \ref{sec:nw_k_choose} provides recommendations for choosing (or tuning) the parameters of the \texttt{mtNUFFT} periodogram and other practical considerations when using multitapering for time-series analysis.

\section{Power Spectrum Estimation in Asteroseismology} \label{sec:multitaper_math}

An important statistical problem in asteroseismology is the detection of oscillation signals given discrete time-series data over a finite time interval. To demonstrate the challenges underlying this problem, in this section we focus on analyzing a light curve for the red giant Kepler-91, shown in Figure \ref{fig:kepler_time}. This analysis draws inspiration from and builds upon the example in \cite{springford_2020}. We refer the reader to this paper for information on the pre-processing of the Kepler-91 light curve.

Figure \ref{fig:kepler_time} shows that the timestamps of the Kepler light curve are unevenly spaced and that long time gaps are present \citep[see][]{kallinger_2014}. This leads us to the first time-series analysis problem in asteroseismology, \textit{irregular sampling}, which we tackle in Section \ref{subsec:sampling}.

Section \ref{subsec:multitaper} discusses the statistical problems of \textit{bias and inconsistency} of the LS, and how the Thomson multitaper approach \citep{thomson_1982} tackles both these problems. In Section \ref{subsubsec:mtnufft}, we introduce \texttt{mtNUFFT}, the extension of multitapering to quasi-regular time sampling.

Finally, we use a simulated Kepler light curve in Section \ref{subsec:simulation_modes} to show that \texttt{mtNUFFT} allows better p-mode modeling than the LS.

\begin{deluxetable}{cl}
    \tablecolumns{2}
    \tablehead{
    \colhead{Symbol} &
    \colhead{Description}
    }
    \tablecaption{Mathematical Notation}
    \label{tab:notation}
    \startdata
        $n$ & sample index in time-series\\
        $\mathbf{x} = \{x_n\}$ & vector of evenly or unevenly-sampled time-series\\
        $\Delta t$ & sampling interval for evenly-sampled $\mathbf{x}$\\
        $\mathbf{t} = \{t_n\}$ & vector of timestamps for unevenly-sampled $\mathbf{x}$\\
        $N$ & sample size of $\mathbf{x}$\\
        $T$ & time duration of $\mathbf{x}$\\
        $\overline{\Delta t}$ & mean sampling interval for unevenly-sampled $\mathbf{x}$\\
        $M$ & zero-padded length of $\mathbf{x}$\\
        $f$ & frequency\\
        $f_\mathrm{Nq}$ & Nyquist frequency\\
        $f_\mathcal{R}$ & Rayleigh resolution\\
        $\tau_\mathrm{LS}$ & time-offset of LS periodogram\\
        $\mathcal{FT}_\mathbf{x}(f)$ & Fourier transform of $\mathbf{x}$\\
        $S(f)$ & true power spectrum underlying $\mathbf{x}$\\
        $\hat S^{(\mathrm{type})}(f)$ & power spectrum estimate of a given type \\
        $W, NW$ & bandwidth, time-bandwidth product\\
        $K$ & number of tapers $\leq 2NW-1$\\
        $k$ & index (order) of taper\\
        $\mathbf{v}(N, W)$ & $K \times N$ matrix of evenly-sampled tapers [$v_{k, n}$]\\
        $\mathbf{v}^{\star}(N, W)$ & $K \times N$ matrix of tapers interpolated to $\mathbf{t}$\\
        $\lambda_k(N, W)$ & eigenvalue of taper $k$\\
        $U_k(N, W; f)$ & Fourier transform of taper $k$ (eigenfunction)\\
        $y_k(f)$ & eigencoefficient of taper $\mathbf{v}_{k}$\\
        $\hat S_k(f)$ & single-tapered power spectrum estimate of order $k$\\
        $d_k(f)$ & adaptive weight of $\hat S_k(f)$\\
        $\hat{S}^{(\mathrm{mt})}(f)$ & multitaper power spectrum estimate\\
        $\hat{S}^{(\mathrm{mt})}_{\setminus j}(f)$ & delete-one [$j$] multitaper power spectrum estimate\\
        $M(\bm{\theta}, \nu)$ & model power spectrum [parameters $\bm{\theta}$, frequency $\nu$]\\
    \enddata
    \tablecomments{We use the above mathematical notation in this paper. Note that we use $\nu$ for model frequency (and $\nu_{nl}$ for stellar modes) instead of $f$ to distinguish between data and theory.}
\end{deluxetable}

\subsection{Sampling of Time-Series Data} \label{subsec:sampling}
The irregularity of Kepler time-series and other space-based observations makes power spectrum estimation in asteroseismology challenging. The statistical behavior of power spectrum estimators in the regularly-sampled case is well understood, making detection of periodic signals in time-series reliable. One such estimator with the simplest statistical behaviour is the \textit{classical periodogram} \citep{schuster_1898}. This estimator is commonly used and is given by

\begin{equation}\label{eq:classicalp}
    \hat S^{(\mathrm{P})}(f) = \frac{1}{N}\left|\sum_{n=0}^{N-1} x_n e^{-i 2\pi f n}\right|^2
\end{equation} where $\mathbf{x} = \{x_n \mid n = 0,...,N-1 \}$ is a zero-mean (strong or weak) stationary\footnote{Statistics underlying a stationary process do not evolve over time.} time-series with sampling $\Delta t = 1$. The summation in Equation~\eqref{eq:classicalp} is the discrete Fourier Transform (DFT) of $\mathbf{x}$, which we denote as $\mathcal{FT}_\mathbf{x}(f)$. By exploiting symmetries in the DFT terms, the Fast Fourier Transform (FFT) algorithm \citep{cooley_1965} can efficiently and accurately compute $\mathcal{FT}_\mathbf{x}(f)$ at $N/2$ regularly-spaced frequencies 
\begin{equation}\label{eq:f_n}
    f_i = i/N \;\; \mathrm{for} \, i=0, 1, \dotsc, \mathrm{floor}(N/2).
\end{equation} Here, the floor function transforms $N/2 \in \mathbb{R}$ into the largest integer $\leq N/2$. These frequencies are equivalent to a \textit{principle frequency domain} of $[-\frac{1}{2}, \frac{1}{2})$. Here, $\frac{1}{2}$ is the Nyquist frequency, i.e., the largest frequency we can completely recover (without aliasing), and $\frac{1}{N}$ is the Rayleigh resolution, i.e., the frequency separation between two just-resolved signals. For any sampling $\Delta t$, the Nyquist frequency is given by
\begin{equation}\label{eq:Nyquist}
    f_\mathrm{Nq} = \frac{1}{2 \Delta t}
\end{equation} whereas the Rayleigh resolution is given by 

\begin{equation}\label{eq:Rayleigh}
    f_\mathcal{R} = \frac{1}{N \Delta t} = \frac{1}{T}.
\end{equation}

The FFT algorithm is orders-of magnitude faster than its ``slow" counterpart. It is most efficient when $N$ is a power of 2, and hence the time-series data $\mathbf{x}$ is \textit{zero padded} to length $M \ge N$, where $M$ satisfies the power of 2 condition. Zero padding by at least a factor of 2 ($M \ge 2N$) can also help circumvent circular correlations. Such a zero-padded version of FFT results in a finer frequency grid as the spacing reduces from $1/N$ to $1/M$. There are many other reasons for zero-padding, and we expand upon some of them using the multitaper F-test in Patil et al. (2024b).

Note that while zero-padding reduces the frequency spacing of the periodogram estimate, the underlying frequency resolution is still limited to the Rayleigh resolution.

While the classical periodogram definition generalizes to irregularly-sampled time-series, its statistical behavior does not directly translate to it. Therefore, certain modifications are necessary which we explore in the following section.

\subsubsection{How to Handle Irregular Sampling?}\label{subsubsec:LS}
The classical periodogram in the regular sampling case has well-defined statistical properties. For example, the periodogram of an evenly-sampled Gaussian noise process has a $\chi^2$ distribution with 2 degrees of freedom ($\chi_2^2$)
\citep{schuster_1898}. This attribute allows us to analyze the presence of spurious peaks in the power spectrum estimates. However, the simple statistical properties of the classical periodogram do not hold in the irregular sampling case, i.e., one cannot define the periodogram distributions analytically. \cite{scargle_1982} tackle this issue by modifying the classical periodogram to the LS periodogram for irregular time sampling. The LS estimator is given by

\begin{figure*}[t]
    \centering
    \includegraphics[width=\linewidth]{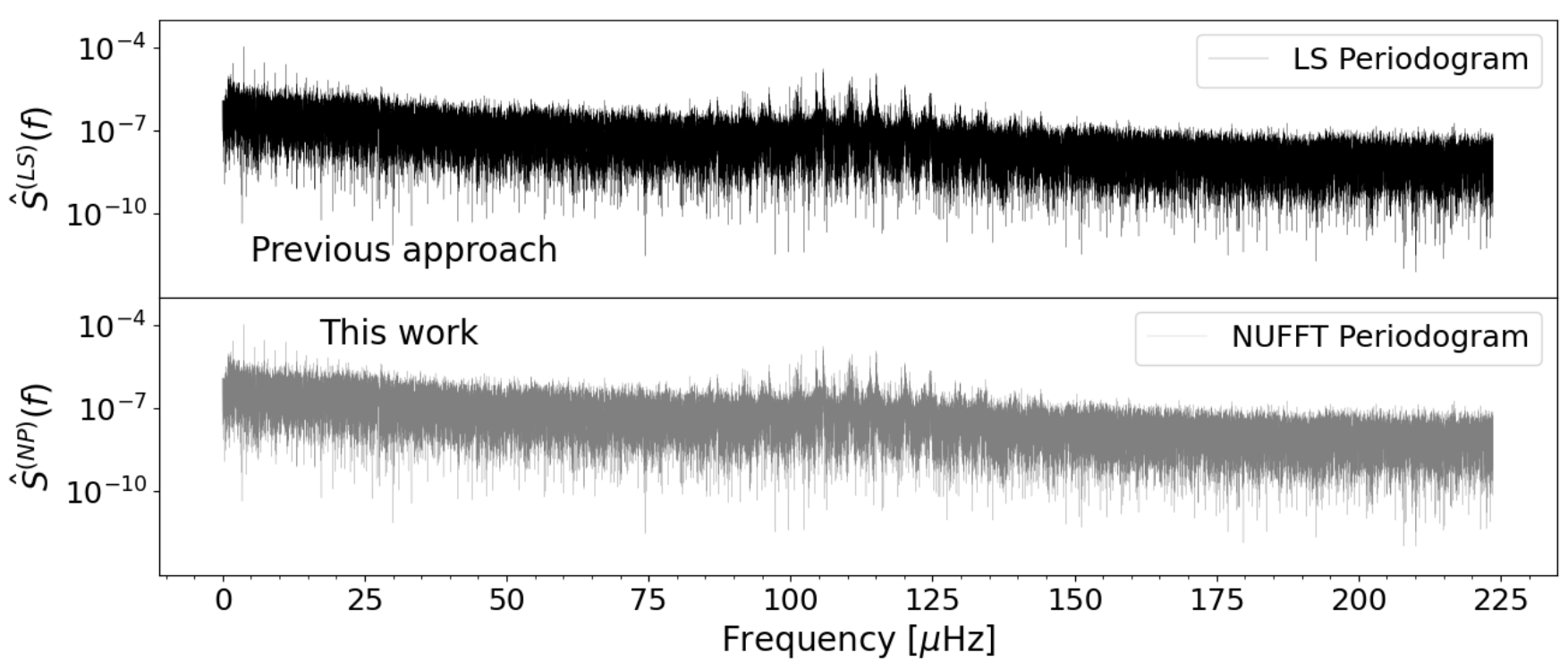}
    \caption{Comparison between the LS periodogram and the \texttt{NUFFT} periodogram of the Kepler-91 time-series. This illustrates that the \texttt{NUFFT} periodogram we introduce in Section \ref{subsubsec:quasi_periodic_nufft} behaves similar to the LS periodogram in the case of quasi-evenly-sampled time-series with gaps.}
    \label{fig:nufft_vs_ls}
\end{figure*}

\begin{multline}\label{eq:LS}
    \hat S^{(\mathrm{LS})}(f) = \frac{1}{2}\frac{ \left\{ \sum\limits_{n=0}^{N-1} x_n cos \left[2 \pi f (t_n - \tau_\mathrm{LS})\right]\right\}^2}{\sum\limits_{n=0}^{N-1} cos^2 \left[2 \pi f (t_n - \tau_\mathrm{LS})\right]}\\
    + \frac{1}{2}\frac{\left\{ \sum\limits_{n=0}^{N-1} x_n sin \left[2 \pi f (t_n - \tau_\mathrm{LS})\right]\right\}^2}{\sum\limits_{n=0}^{N-1} sin^2 \left[2 \pi f (t_n - \tau_\mathrm{LS})\right]}
\end{multline} where $\mathbf{x} = \{x_n\}$ corresponding to timestamps $\mathbf{t} = \{t_n \mid n = 0,...,N-1\}$ is an irregularly-sampled time-series. $\tau_\mathrm{LS}$ is the time-offset given by
\begin{equation} \label{eq:tau_LS}
    \tan \left( 2 f \tau_\mathrm{LS} \right) = \frac{\sum\limits_{n=0}^{N-1} sin \left( 4 \pi f t_n \right) }{\sum\limits_{n=0}^{N-1} cos \left( 4 \pi f t_n \right) }
\end{equation} that makes the periodogram invariant to time-shifts. The distribution of this modified periodogram is equivalent to the classical periodogram.

The LS periodogram was designed to detect a single periodic signal embedded in normally distributed independent noise \citep{scargle_1982}. It is essentially a Fourier analysis method that is statistically equivalent to performing least-squares fitting to sinusoidal waves \citep{lomb_1976}, which can be shown using Equation \eqref{eq:LS}. We refer the reader to \cite{vanderplas_2018} for an in-depth review of the LS periodogram estimator.

\subsubsection{Quasi-Regular vs Irregular Sampling}\label{subsubsec:quasi_periodic_nufft}
Given the irregular time sampling of space-based light-curves such as those from Kepler, the LS periodogram is the preferred power spectrum estimator. 
However, between the long downlink gaps, the time steps are almost identical and the light curves can be treated as quasi-evenly sampled \citep{fodor_2000, smith_2012, pires_2015}. In this case, the statistical properties of the classical periodogram should hold to some degree. Taking advantage of this, we implement a periodogram for quasi-regular sampling using the Non-Uniform or non-equispaced FFT (\texttt{NUFFT}) \citep{keiner_2009, barnett_2018}. Essentially, we directly generalize the classical periodogram to the irregular sampling case as

\begin{equation}\label{eq:NUFFT_period}
    \hat S^{(\mathrm{NP})}(f) = \frac{1}{N}\left|\sum_{n=0}^{N-1} x_n e^{-i 2\pi f t_n}\right|^2
\end{equation} and compute the non-uniform DFT in the definition using the adjoint \texttt{NUFFT}. This \texttt{NUFFT} takes unevenly-sampled time series and computes corresponding Fourier series coefficients on a uniform frequency grid as follows

\begin{equation}
    c_i = \sum_{n=0}^{N-1} x_n e^{-i 2\pi f_i t_n}
\end{equation} where $c_i$ are the complex coefficients for frequency $f_i$ defined in Equation \eqref{eq:f_n}. One can apply zero-padding to the adjoint \texttt{NUFFT} in the same way as to the FFT.

We can think of the \texttt{NUFFT} periodogram as a simpler version of the LS; instead of using the adjoint \texttt{NUFFT} directly to compute Equation \eqref{eq:NUFFT_period}, the LS uses the transform to compute the modified components in Equation \eqref{eq:LS}. This simplicity makes the \texttt{NUFFT} periodogram slightly more efficient than the LS. In addition to efficiency, the direct usage of \texttt{NUFFT} offers complex Fourier coefficients with phase information that one can leverage to distinguish between transient (damped) and strictly periodic (undamped) oscillatory modes in asteroseismology (refer to Patil et al. 2024b for more details). We therefore suggest using the \texttt{NUFFT} periodogram for quasi-regular sampling due to its simplicity and the phase information it provides.

Figure \ref{fig:nufft_vs_ls} compares the \texttt{NUFFT} periodogram with the LS periodogram for the Kepler-91 time-series. We use the adjoint \texttt{NUFFT} from the \texttt{FINUFFT} package\footnote{\url{https://github.com/flatironinstitute/finufft}} \citep{barnett_2019, barnett_2021} and the default \texttt{astropy} LS implementation for computing the two periodograms. Both have a frequency grid with an oversampling factor of 5. A comparison between the two power spectrum estimates shows that, excluding some random variations across the two periodograms that follow their distribution properties, the two estimates agree with each other. They are both able to extract the comb-like p-mode structure around the frequency of 115 $\mu$Hz. Thus, in practice, the modification of the classical periodogram to LS is not necessary for the quasi-regular sampling of Kepler time-series and we can directly apply the \texttt{NUFFT} periodogram. 

There are subtle differences in the amplitudes of the low frequency signals on top of the granulation background in Figure \ref{fig:nufft_vs_ls}, which could be due to the different ways in which the two estimators detect periodic components. These differences should scale with the irregularity of the time-samples. In theory, the LS works better for highly irregular or random time samples, whereas \texttt{NUFFT} and LS should work similarly for quasi-even sampling.

\begin{figure*}[ht]
    \centering
    \includegraphics[width=\linewidth]{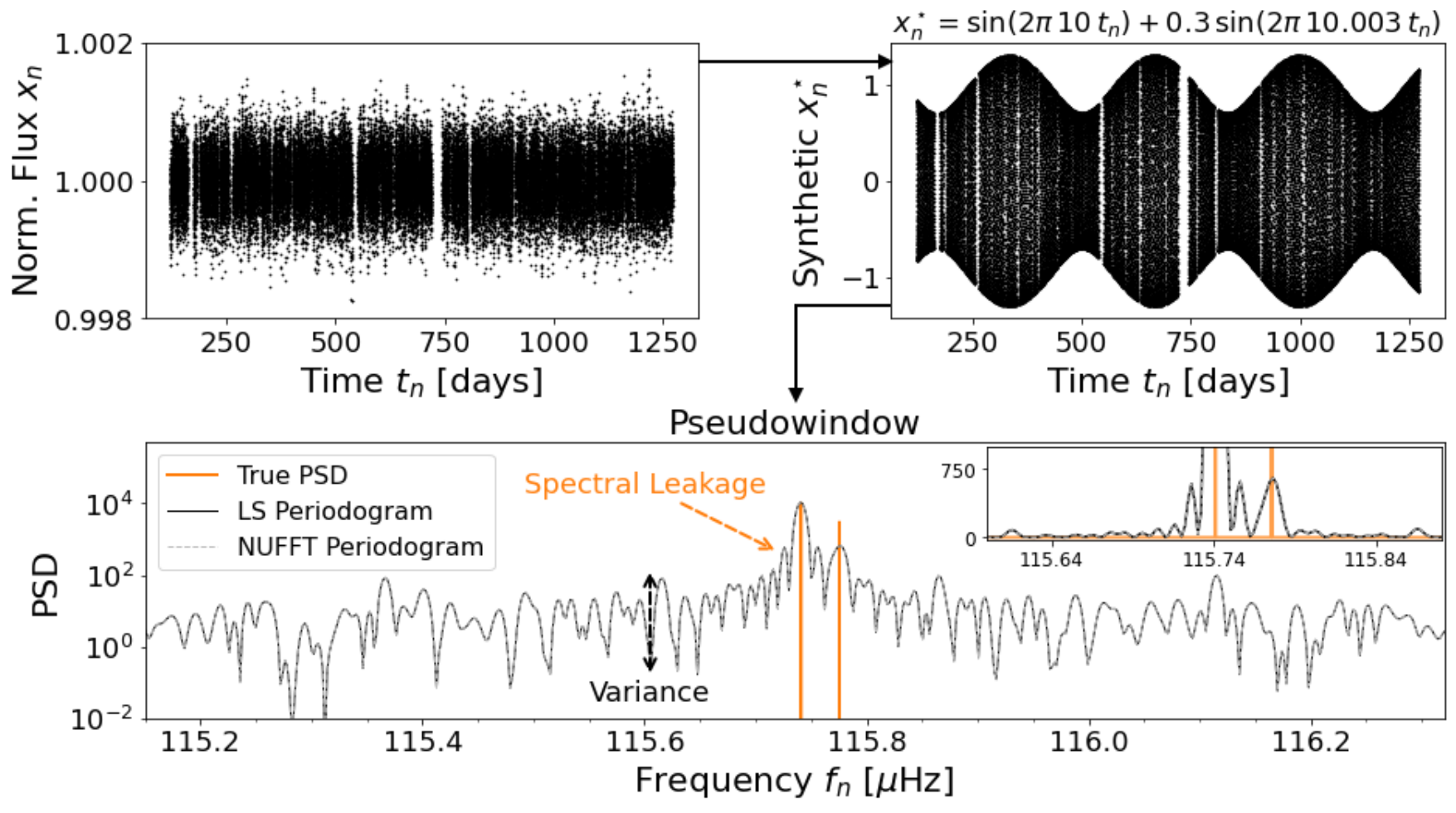}
    \caption{Power spectrum analysis of the pseudowindow generated using the irregular sampling times of the Kepler-91 light curve shown in the top left panel. The top right panel displays the synthetic light curve composed of two sinusoidal signals sampled at the times $t_n$ of the Kepler-91 series. The bottom panel shows a zoomed-in version of the LS (black) and \texttt{NUFFT} (grey) periodograms of the synthetic light curve $x^\star(t_n)$ as compared to the true PSD (orange). The two estimates of the true spectrum are nearly identical, and both show inconsistency and spectral leakage compared to the true PSD. Note that the bottom panel is in log scale, whereas its inset is in linear scale to better view the difference between the signals and the noise.}
    \label{fig:psuedo_windows}
\end{figure*}

\subsection{Statistical issues with the Periodogram}\label{subsec:multitaper}
While the LS periodogram solves the problem of detecting a periodic signal in irregularly-sampled data and has simple statistical behaviour, it suffers from the problems of inconsistency and spectral leakage that are inherent to the analysis of a finite, discrete, and noisy time-series. They are as follows:

\begin{enumerate}
    \item \textit{Inconsistency}:
    An inconsistent estimator is one whose variance does not tend to zero as the sample size $N \to \infty$. The variance of the estimator is high even for data with high SNR and it does not reduce with increasing $N$. For e.g., the LS periodogram of a Gaussian noise process is exponentially ($\chi^2_2$) distributed with large variance. The variance also does not reduce as $N$ increases because the number of frequencies recovered by the estimate, given by $N/2$ as in Equation \eqref{eq:f_n}, proportionally increases.
    \item \textit{Spectral leakage}:
    Spectral leakage refers to the manifestation of power belonging to frequency $f$ at a far-away frequency $f + \Delta f$. Several sources of leakage are known to affect power spectrum estimates. The finite time interval of time-series observations represents a rectangular window and leads to side lobes that cause leakage to nearby frequencies. Here, a side lobe refers to any peak in the Fourier transform of a window function that is outside the main peak (refer to the side lobe labeled as ``Spectral Leakage" in Figure \ref{fig:psuedo_windows}). In contrast, the discreteness of the time-series causes leakage to distant frequencies. Thus, leakage can lead to badly biased power spectrum estimates, especially when the sample size $N$ is small. 
\end{enumerate}

The classical periodogram faces the same issues albeit with a smaller degree of spectral leakage. We can analytically define the power spectral window function (the frequency response of a time-domain window) for evenly-sampled data which completely describes the spectral leakage properties of the periodogram. In contrast, the spectral leakage of the LS periodogram does not have a simple analytical definition. It depends on the exact time-sampling structure, is frequency-specific, and is often worse than that of the periodogram. 

We can visualize the spectral leakage properties of the LS and NUFFT periodograms by investigating the \textit{pseudowindow} in Figure \ref{fig:psuedo_windows}. A pseudowindow is the response of a power spectrum estimator to a pure sinusoidal signal of a given frequency with the same sampling as the time-series of interest. We create a signal with two sinusoids $x^\star(t) = \sin(2\pi 10 t) + 0.3 \sin(2\pi 10.003 t)$ of frequencies $10$ and $10.003$ cycles/day (or $115.74$ and $115.78 \mu$Hz) respectively and sample them at the times of the Kepler-91 series. The bottom panel of Figure \ref{fig:psuedo_windows} displays the true Power Spectral Density (PSD) of the synthetic light curve that is given by two delta functions at the frequencies of the sinusoids with heights equal to the sinusoid amplitudes, plus the PSD estimates returned by the LS (black line) and NUFFT (grey dashed line) periodograms. Particularly, we see that the leakage of power from the two sinusoid frequencies results in spurious peaks in their vicinity. These peaks can lead to false discoveries when analyzing Kepler time-series (refer to \citealt{vanderplas_2018} for more details).

\subsubsection{How does the Multitaper Power Spectrum Estimate help?}\label{subsubsec:mt_general_solution}
As discussed earlier, the motive in \cite{scargle_1982} was to detect a strictly periodic (sinusoidal) component embedded in a Gaussian noise process. Several applications in astronomy show that, in addition to sinusoids, LS works well at estimating the frequencies of non-sinusoidal (or quasi-)periodic signals in time-series \citep[e.g.,][]{baliunas_1995, robertson_2012, reinhold_2013}. However, the spectral leakage properties of the LS estimator are poor, especially if the underlying power spectrum is not of the type envisioned. Thus, \cite{scargle_1982} suggests computing the LS periodogram on tapered time-series data to mitigate spectral leakage \citep{brillinger_1981}.

Tapering a time-series reduces spectral leakage, but there is a trade-off between bias control and variance reduction. Essentially, a taper down weights the edges of a time-series to reduce spectral leakage, and therefore bias, in the power spectrum estimate. However, tapering also results in an increase in the variance of the estimate. Thus, the appropriate taper choice depends on the way the bias/variance trade-off affects the science goals. 

Instead of using a single-tapered power spectrum estimate, \cite{thomson_1982} develop the multitaper estimate which uses DPSS \citep{slepian_1978} as tapers to optimally reduce spectral leakage along with variance. The tapers are orthogonal to each other and hence provide independent estimates of the power spectrum, which are averaged to minimize variance as follows

\begin{equation}\label{eq:mt_spec}
    \hat{S}^{(\mathrm{mt})}(f) = \frac{1}{K} \sum_{k=0}^{K-1} \hat{S}_k(f).
\end{equation} Here, $\hat{S}_k(f)$ are single-tapered, independent power spectrum estimates that are averaged to get the multitaper estimate.

Thus, both spectral leakage and inconsistency are tackled by the multitaper estimate, and this makes it an improvement over the classical periodogram in the even sampling case as well as the LS periodogram in the uneven sampling case. While the multitaper estimate was originally developed for a regularly-sampled time-series, a multitaper version of the LS periodogram was recently developed for irregular sampling \citep[\texttt{mtLS};][]{springford_2020}. We discuss the multitaper estimate for regularly-sampled times in Section \ref{subsubsec:mt}, and introduce our new \texttt{mtNUFFT} method, an extension of the \texttt{mtLS} periodogram, to quasi-regular time sampling in Section \ref{subsubsec:mtnufft}.

\begin{figure*}[t]
    \centering
    \includegraphics[width=\linewidth]{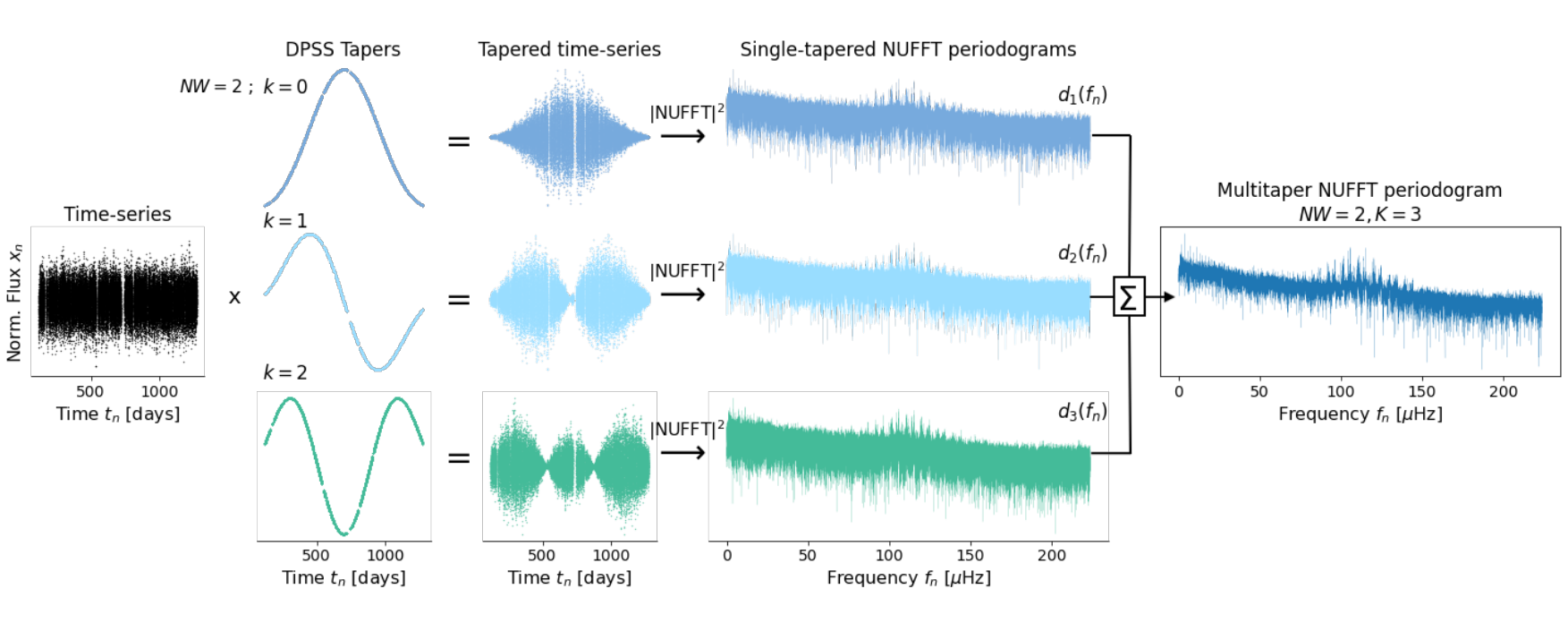}
    \caption{Schematic diagram illustrating the estimation of the \texttt{mtNUFFT} periodogram described in Section \ref{subsubsec:mtnufft}. The left panel shows the Kepler-91 time-series for which we compute the power spectrum estimate. The middle panels show three DPSS or Slepian tapers with time-bandwidth product $NW=2$ and order $k=0, 1, 2$ (number of tapers $K=2NW-1$), their corresponding tapered time-series, and the single-tapered NUFFT periodograms. The rightmost panel shows the multi-tapered NUFFT or \texttt{mtNUFFT} periodogram that is constructed by averaging the three single-tapered estimates with adaptive weights $d_k(f)$.}
    \label{fig:mt_process}
\end{figure*}

\subsubsection{Multitaper Power Spectrum Estimate for Regular Sampling}\label{subsubsec:mt}
\cite{thomson_1982} develop the multitaper estimate as an approximate solution of the fundamental integral equation of power spectrum estimation by performing a ``local" eigenfunction expansion. We refer the reader to \cite{thomson_1982} and \cite{percival_1993} for more details on the mathematical theory.

The multitaper estimate $\hat{S}^{(\mathrm{mt})}(f)$ of the true power spectrum $S(f)$ underlying an evenly-sampled time-series $\mathbf{x}$ is an average of $k=0,1,\dotsc,K-1$ independent power spectrum estimates $\hat{S}_k(f)$ computed using orthonormal DPSS tapers $\mathbf{v}_k(N, W)$ [refer to Equation \eqref{eq:mt_spec}]. These tapers are the same length as the time-series, indexed as $v_{k, n}(N, W)$ for $n=0, 1, \dotsc, N-1$ \citep[following the notation in][]{slepian_1978}. Each taper $\mathbf{v}_k(N, W)$ corresponds to an eigenfunction $U_k(N, W; f)$ (also called discrete prolate spheroidal wave function), which is the Fourier transform of the taper. These eigenfunctions satisfy the following equation

\begin{multline}
         \int_{-W}^{W} \left[\frac{\sin N \pi (f - f')}{\sin \pi (f - f')}\right] U_k(N, W; f') df' \\
         = \lambda_k(N, W) \cdot U_k(N, W; f)
\end{multline} where $\lambda_k(N, W)$ are the eigenvalues. The first factor on the left-hand side of this equation is the Dirichlet kernel that represents the spectral leakage due to the use of a rectangular window (finite length time-series). Thus, if the eigenvalue $\lambda_k(N, W)$ corresponding to an eigenfunction $U_k(N, W; f)$ is close to 1, most of the power at frequency $f$ is concentrated in the range $(f-W, f+W)$, where $W$ is the bandwidth that is generally on the order of $1/N$. The fraction of the total power at frequency $f$ concentrated in this range is defined as \textit{in-band fractional power concentration}.

The zeroth-order eigenfunction $U_0(N, W; f)$ [or taper $\mathbf{v}_0(N, W)$] has the greatest in-band fractional power concentration, which reduces as the order $k$ increases. We can show this through the ordering of the eigenvalues $\lambda_k$
\begin{equation}
    1 > \lambda_0 > \lambda_1 > \dotsc > \lambda_{K-1} > 0,
\end{equation} which represent the concentration of tapers $\mathbf{v}_{k}(N, W)$. Thus, the bandwidth $W$ is an important parameter that defines the behaviour of the tapered power spectrum estimates, and is chosen by the user based on their science goals.

Note that one generally quotes the time-bandwidth product $NW$\footnote{Generally, $C_\mathcal{R} = N \times W = T \times B$ is the time-bandwidth product, where $B$ is the bandwidth. For a time-series with unit sampling $\Delta t = 1$, $N \equiv T$ and $W$ is the bandwidth.} (as opposed to just the bandwidth $W$) for the multitaper power spectrum estimator. This is because $NW$ is the bandwidth in units of Rayleigh resolution $f_\mathcal{R}$, the frequency separation between two just-resolved signals (Equation \ref{eq:Rayleigh}). The number of tapers that have large in-band fractional power concentration (or minimal out-of-band spectral leakage) is then given by $K \lessapprox \mathrm{floor}(2 NW)$. 

A multitaper estimator with $NW=1$ has a bandwidth equivalent to the Rayleigh resolution, but it only allows the use of $K \approx 1$ taper to avoid badly-biased estimates. In general, $NW$ must be $>1$ and is chosen based on a compromise between frequency resolution and statistical stability (low bias and variance) of the multitaper estimator. For example, the bandwidth for $NW=4$ is four times the Rayleigh resolution, but the number of tapers $K$ we can safely use is $K \approx 7$.

We provide several tips for choosing both $NW$ and $K$ in the Appendix \ref{sec:nw_k_choose}. A rule of thumb is to use $NW=4$ if one knows that the power spectrum has a large dynamic range, i.e., the ratio between maximum and minimum values of $S(f)$ is high, and $K = \mathrm{floor}(2NW) - 1$ tapers. 

We show three DPSS tapers with $NW=2$ and order $k=0, 1, 2$ in Figure \ref{fig:mt_process}. Note that the tapers in this figure are unevenly-sampled, and are used to compute the \texttt{mtNUFFT} periodogram described later.

\textit{Eigencoefficients} corresponding to each taper are defined by the following DFT

\begin{equation}\label{eq:mt_eigen}
    y_k(f) = \sum_{n=0}^{N-1} v_{k, n} x(t) e^{-i 2 \pi f n}
\end{equation} which we can compute using the (zero-padded) FFT algorithm (refer to Section \ref{subsec:sampling}).

\begin{figure*}[t]
    \centering
    \includegraphics[width=0.9\linewidth]{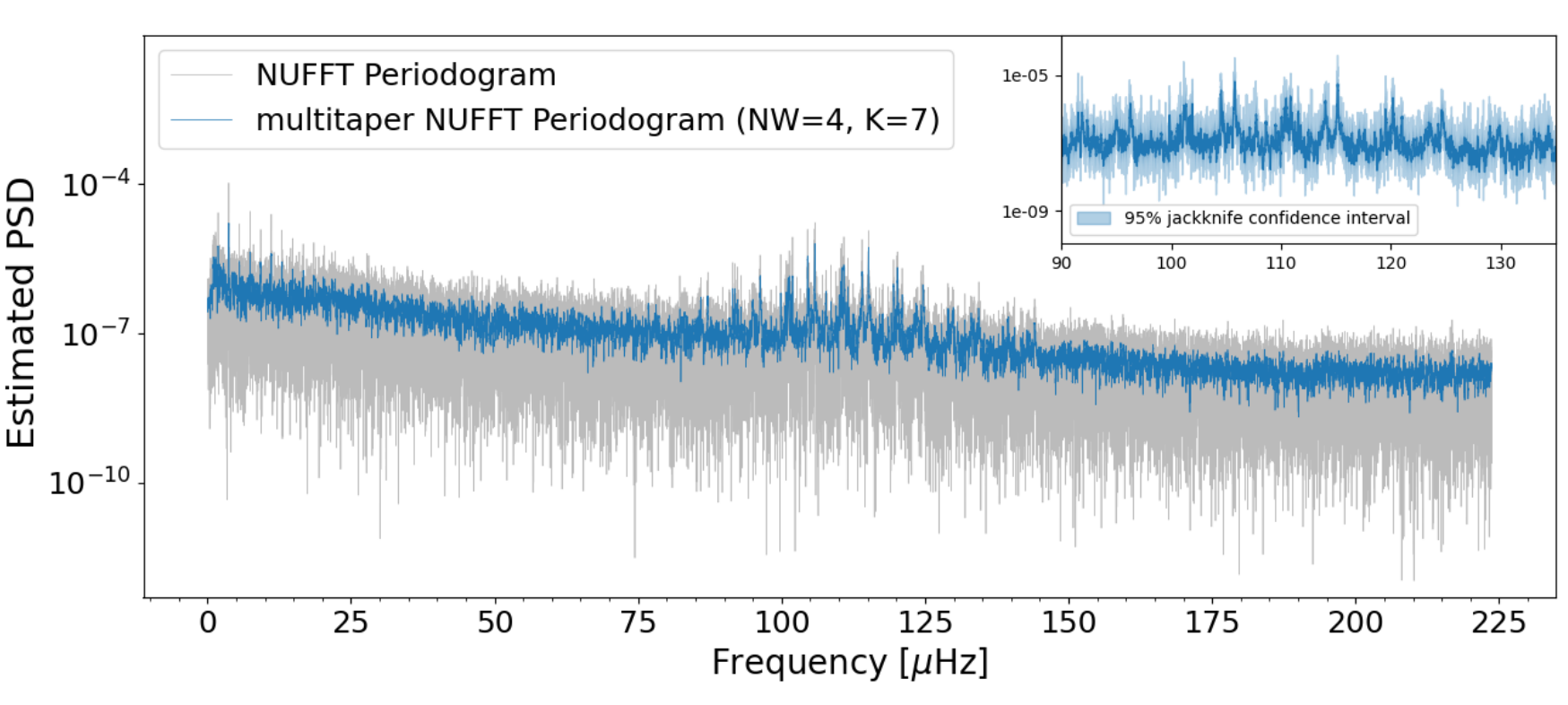}
    \caption{Comparison between the NUFFT and the \texttt{mtNUFFT} periodograms of the Kepler-91 time-series. The parameters of the multitaper periodograms are $NW=4$ and $K=7$. The multitaper NUFFT periodogram in blue has smaller variance compared to the un-tapered counterpart in grey. The inset shows the zoomed-in \texttt{mtNUFFT} periodogram along with its 95\% jackknife confidence interval.}
    \label{fig:mtnufft_vs_mtLS}
\end{figure*}

We can then compute the multitaper power spectrum estimate as shown in Equation \eqref{eq:mt_spec}, given that 
\begin{equation}\label{eq:mean_eigen}
    \hat S_k(f) = \left|y_k(f)\right|^2
\end{equation} is the $k$th eigenspectrum.

Instead of taking an average, we can weight each eigencoefficient $y_k(f)$ using an iterative \textit{adaptive weighting} procedure to improve bias properties. Higher order tapers lead to progressively more spectral leakage and therefore are downweighted using adaptive weights $d_k(f)$ to obtain a less biased multitaper power spectrum estimate (refer to \citealt{thomson_1982} for more details). The schematic diagram in Figure \ref{fig:mt_process} illustrates the above described steps to compute multitaper power spectrum estimates.

We can also estimate \textit{confidence intervals} on the multitaper power spectrum estimate by \textit{jackknifing over tapers} \citep{thomson_1991}. Jackknifing proceeds as follows. First, one computes delete-one power spectrum estimates $\hat{S}^{(\mathrm{mt})}_{\setminus j}(f)$ by omitting the $j$th eigencoefficient from Equation \eqref{eq:mt_spec}, i.e.,

\begin{equation}
    \hat{S}^{(\mathrm{mt})}_{\setminus j}(f) = \frac{1}{K-1} \sum_{k=0, k \neq j}^{K-1} \left|y_k(f)\right|^2.
\end{equation} We denote the average of these estimates as
\begin{equation}
\hat{S}^{(\mathrm{mt})}_{\setminus \bullet}(f) = \frac{1}{K} \sum_{j=0}^{K-1}  \hat{S}^{(\mathrm{mt})}_{\setminus j}(f).
\end{equation}
Then, the jackknife variance of $\hat{S}^{(\mathrm{mt})}(f)$ is given by

\begin{equation}\label{eq:jk_var}
\widehat{\mathrm{Var}}_J(f) = \frac{(K-1)^2}{K(K - \frac{1}{2})} \sum_{j=0}^{K-1} [\ln \hat{S}^{(\mathrm{mt})}_{\setminus j}(f) - \ln \hat{S}^{(\mathrm{mt})}_{\setminus \bullet}(f)]^2.
\end{equation} The 95\% jackknife confidence intervals ($\alpha = 0.05$ significance level) are then constructed using the Student's t-distribution with $K-1$ degrees of freedom. Refer to Equation (2.49) in \cite{thomson_1991} for more details.

\cite{thomson_1994} show that Equation \eqref{eq:jk_var} using the logarithmic power spectrum $\ln \hat{S}^{(\mathrm{mt})}_{\setminus j}(f)$ is a more accurate estimator of $\mathrm{Var}\{S(f)\}$ than the direct variance obtained from individual eigenspectra $\hat S_k(f)$ \citep{thomson_1991}. Since the jackknife variance estimate is distribution-free, it is valid even when the data are non-stationary and/or the observer cannot assume the data follow a particular analytical distribution (e.g. Gaussian or chi2). It is also an efficient\footnote{Statistical efficiency of an estimator determines how well it utilizes available data to provide a precise estimate.} variance estimator as compared to the direct variance, i.e., the estimate starts to approach the true variance even for small $K$.

We can understand this efficiency by looking at the example of Gaussian stationary data. For such data, the distribution of $\hat S_k(f)$ is $\chi_2^2$, i.e., exponential. The highest probability value of $\chi_2^2$ is zero and its logarithm $\ln \hat S_k(f)$ is distributed such that it has a heavy tail. Thus, this distribution inflates the direct variance estimate. On the other hand, the jackknifed $\hat{S}^{(\mathrm{mt})}_{\setminus j}(f)$ have $\chi_{2K-2}^2$ distributions that approach normality with increasing $K$, and provide conservative variance estimates. $K$ does not need to be large for this to be true. For e.g., the variance of a $\ln \chi_{2}^2$ distributed estimate is 1.6449 whereas it is 0.6449 for $\chi_{4}^2$ \citep[shown in][]{thomson_2007}. We refer the reader to Section \ref{subsec:simulation_modes} for more details on the behaviour of $\chi_2$ distributions. Figure \ref{fig:mtnufft_vs_mtLS} shows jackknife confidence intervals for the \texttt{mtNUFFT} power spectrum estimate described in Section \ref{subsubsec:mtnufft}.

\begin{figure*}[ht]
    \centering
    \includegraphics[width=\linewidth]{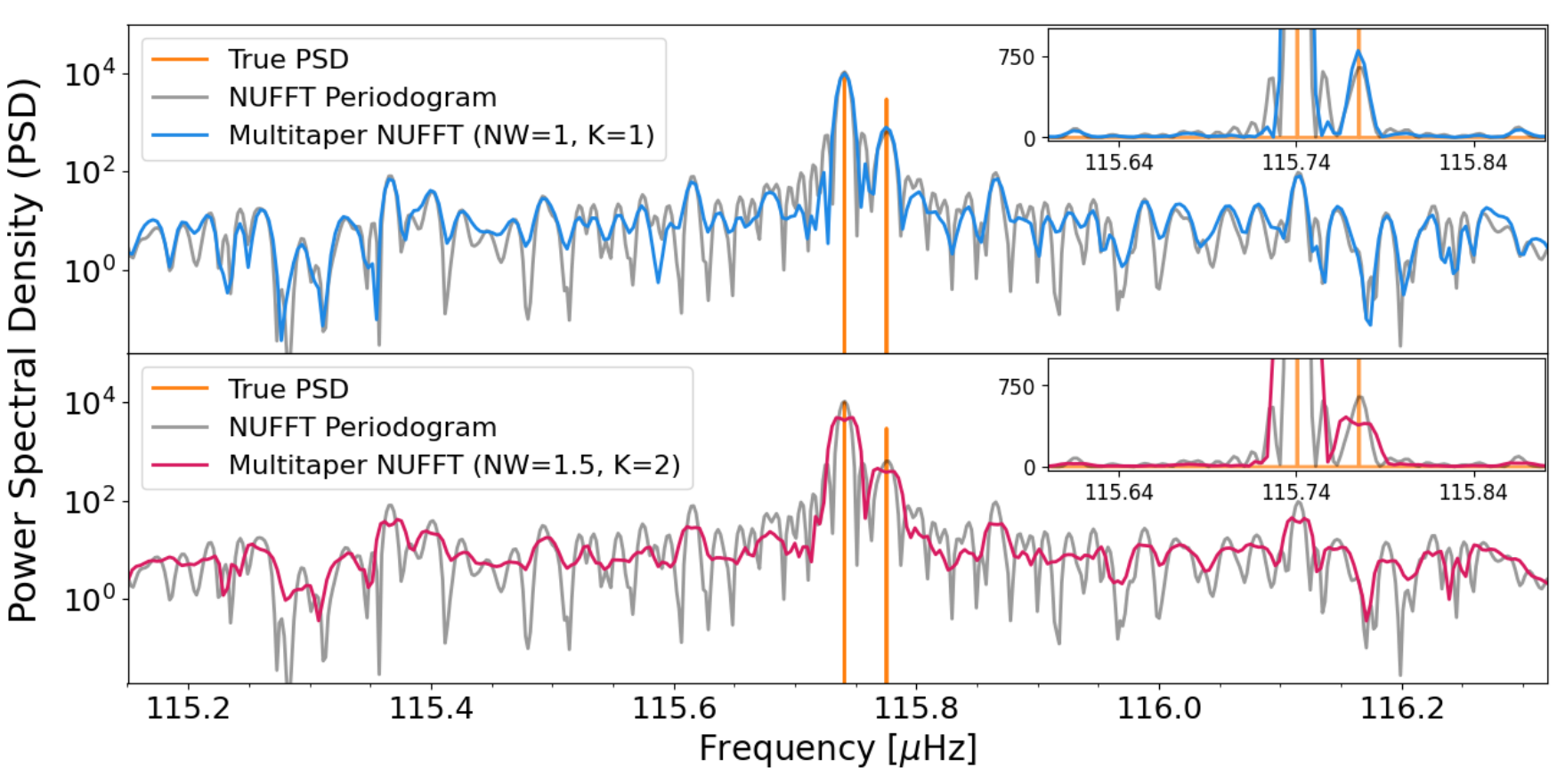}
    \caption{Both panels show the same pseudowindow as that in the bottom panel of Figure \ref{fig:psuedo_windows}, but for the \texttt{NUFFT} and \texttt{mtNUFFT} periodograms. The top panel shows the spectral leakage properties of the \texttt{mtNUFFT} periodogram with $NW=1$ and $K=1$, i.e., the single-tapered power spectrum estimate, in blue, whereas the bottom panel shows the $NW=1.5$, $K=2$ \texttt{mtNUFFT} estimate in dark pink. It is clear that the \texttt{mtNUFFT} estimates have smaller spectral leakage than \texttt{NUFFT}. In the bottom panel, we observe that as $NW$ increases, the variance of the estimate reduces but the frequency resolution worsens.}
    \label{fig:pseudo_window_mt}
\end{figure*}

\subsubsection{Multitaper NUFFT for Quasi-Regular Sampling}\label{subsubsec:mtnufft}
In Section \ref{subsubsec:quasi_periodic_nufft}, we presented the \texttt{NUFFT} periodogram that works well for detecting stellar oscillatory modes in quasi-regularly sampled time-series. We can combine this periodogram with the multitaper statistic to obtain the \texttt{mtNUFFT} periodogram. To compute this periodogram, we use the same procedure as that described in \cite{springford_2020} to compute the \texttt{mtLS} periodogram. We compute DPSS tapers $\mathbf{v}_{k}(N, W)$ of order $k=0,\dotsc,K-1$ on a regular grid with sampling interval $\overline{\Delta t} = T/N$ where $T = t_{n-1} - t_0$. Then, we interpolate these tapers to the uneven sampling times $\mathbf{t}$ using a cubic spline and renormalize them. These interpolated and renormalized tapers are referred to as $\mathbf{v}^{\star}_k(N, W)$. The only modification of the procedure in \cite{springford_2020} is that instead of computing $K$ independent LS periodograms $\hat{S}^{(\mathrm{LS})}_k(f)$ on the tapered time-series $v_{k, n}^{\star} x_n$, we compute the eigencoefficients

\begin{equation}\label{eq:mtnufft}
    y_k(f) = \sum_{n=0}^{N-1} v_{k, n}^{\star} x_n e^{-i 2 \pi f t_n},
\end{equation} using the (zero-padded) adjoint \texttt{NUFFT} to obtain the $\hat{S}_k^{(\mathrm{mt})}(f)$ and $\hat{S}^{(\mathrm{mt})}(f)$ through Equations \eqref{eq:mt_spec} and \eqref{eq:mean_eigen}, respectively. These eigencoefficients are the generalization of Equation \eqref{eq:mt_eigen} to the case of irregular sampling.

The \texttt{mtNUFFT} estimation procedure is shown in Figure \ref{fig:mt_process}. It is important to note that the cubic spline interpolation of DPSS tapers maps the evenly-sampled tapers to the irregularly spaced times $\mathbf{t}$, but does not fully maintain the tapers' optimal in-band power concentration. Taper interpolation introduces out-of-band spectral leakage beyond that of the classical DPSS tapers. 

We show tapers interpolated to the Kepler-91 timestamps in Figure \ref{fig:mt_process}. This figure demonstrates that the tapers retain their shapes after interpolation because Kepler time sampling is quasi-regular with gaps that are not too long. Thus, interpolation in such cases leads to minor increase in spectral leakage compared to that of the classical tapers \citep{springford_2020}. On the other hand, interpolating tapers to highly irregular time samples should produce a power spectral window with large spectral leakage. In this case, one should use the quadratic power spectrum estimator of \cite{bronez_1988} that evaluates generalized DPSS to achieve minimal spectral leakage for irregularly-sampled time-series. However, it comes at the expense of a computationally intensive matrix eigenvalue problem. In comparison, the \texttt{mtNUFFT} periodogram is fast to compute (similar to \texttt{mtLS}) and a significant improvement over LS, which is why we use it in this study.

In Figure \ref{fig:mtnufft_vs_mtLS}, we compare the \texttt{mtNUFFT} periodogram with the \texttt{NUFFT} periodogram. The two power spectrum estimates are on the same frequency grid with an oversampling factor of 5. We extend the jackknife confidence intervals of the multitaper statistic for evenly-sampled time series \citep{thomson_1982} to the \texttt{mtNUFFT}. Figure \ref{fig:mtnufft_vs_mtLS} shows the jackknife confidence interval of the \texttt{mtNUFFT} periodogram of the Kepler-91 time-series.

We note that the canonical statistical properties of the jackknife variance estimator do not directly apply to \texttt{mtNUFFT} because it interpolates DPSS tapers to unevenly-sampled time-series. In general, orthogonality of DPSS tapers is lost upon interpolation, which means that the $\hat{S}^{(\mathrm{mt})}_k(f)$ are not independent and their jackknife variance cannot be estimated \citep[refer to][]{dodson-robinson_2022b}. However, we find that deviation from orthogonality is minor for quasi-regular sampling, e.g., of Kepler time-series. We show this in Appendix \ref{sec:jackknife} by regularly and quasi-regularly sampling a process with a known power spectrum and comparing the jackknife confidence intervals of the two time-series. However, we reiterate that for highly irregular time samples, the jackknife variance estimator is not applicable.

In Figure \ref{fig:pseudo_window_mt}, we use pseudowindows to show that spurious peaks in the \texttt{NUFFT} periodogram (grey) that are not at the frequencies of the true signals (orange) are suppressed in the \texttt{mtNUFFT} periodogram (light blue). Figure \ref{fig:pseudo_window_mt} also shows that as $NW$ increases, the number of tapers one can use to generate the \texttt{mtNUFFT} periodogram with negligible out-of-band spectral leakage also increases ($K=2NW-1$), leading to a power spectrum estimate with reduced variance (dark pink). However, the frequency resolution worsens due to increased \textit{local} bias. Essentially, local bias results from averaging the power spectrum or flattening its structure over the interval $(f-W, f+W)$, which lowers its frequency resolution. We discuss this trade-off, and provide advice in choosing the parameters $NW$ and $K$, in the Appendix \ref{sec:nw_k_choose}. We note that similar trade-offs between spectral leakage, variance, and bandwidth (resolution) apply to some other power spectrum estimators \citep[e.g.,][]{welch_1967}, but the advantage of using the multitaper estimator is that given two of these performance measures, it minimizes the third one \citep[][more details in Section~\ref{sec:discussion}]{bronez_1992}.

In the following section, we use a simulated asteroseismic time-series of a solar-like oscillator to illustrate that we can accurately model p-modes using \texttt{mtNUFFT}, significantly better than LS. These p-modes can then inform the theory of stellar structure and evolution, and allow precise estimates of mass, radius, age, and other fundamental stellar properties. 

\begin{figure*}
    \centering
    \includegraphics[width=\linewidth]{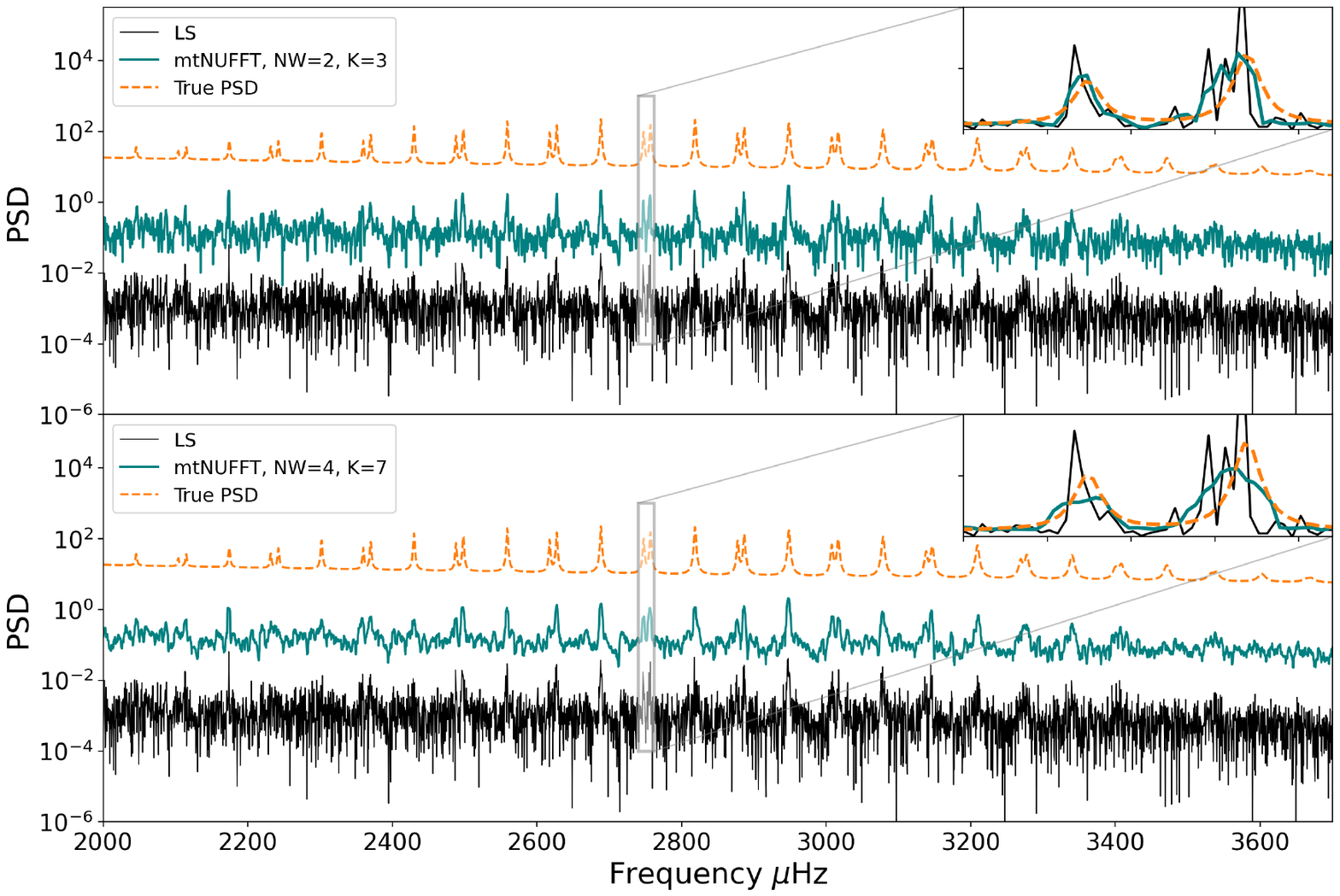}
    \caption{Comparison between LS, \texttt{mtNUFFT}, and the true spectrum used to simulate an asteroseismic time-series (refer to Section \ref{subsec:simulation_modes} for more details of the simulation). The top and bottom panels show the $NW=3, \, K=5$ and $NW=4, \, K=7$ mtNUFFT periodograms respectively. The insets at the top right of the two panels zoom into two p-modes; mtNUFFT is able to estimate the PSD more accurately than the LS by reducing both bias and variance. We also see that the resolution slightly reduces as we increase $NW$, but it does not affect mode estimation in this case.}
    \label{fig:simulation}
\end{figure*}

\subsection{Simulated Time-Series of a Solar-like Oscillator}\label{subsec:simulation_modes}
To illustrate the power spectrum estimation accuracy of the \texttt{mtNUFFT} periodogram, we simulate a light curve of a solar-like oscillator using an asteroseismic power spectrum model of a $1.0\;\mathrm{M}_{\odot}$ star of age $3.99$ Gyr and $Z=0.01$. We use a similar procedure as \cite{ball_2018} to simulate our synthetic (model) power spectrum containing a granulation background and a p-mode envelope of a sum of Lorentzians

\begin{equation}\label{eq:astero_model}
    M(\bm{\theta}, \nu) = b + \sum\limits_{n=1}^{N} \frac{h_n}{1 + \frac{4}{w_n^2} (\nu - \nu_n)^2}.
\end{equation} Here the parameters $\bm{\theta}$ are $T_\mathrm{eff}, \Delta \nu, \nu_\mathrm{max}, \epsilon$ (and more depending on the complexity of the model) which determine the background $b$, and heights $h_n$, widths $w_n$, frequencies $\nu_n$ of the Lorentzian profiles of the $N$ modes. 

$\bm{\theta}$ for a given stellar mass (and age) are computed using the two scaling relations for global asteroseismic properties \citep{kjeldsen_1995}. The following scaling relation connects the large frequency separation $\Delta \nu$ of p-modes of a star to the density (i.e., mass $M$ and radius $R$) of the star

\begin{equation}\label{eq:astero1}
    \frac{\Delta \nu}{\Delta \nu_{\odot}} \simeq \left(\frac{M}{M_{\odot}}\right)^{1/2} \left(\frac{R}{R_{\odot}}\right)^{-3/2}.
\end{equation} Here $\Delta \nu_{\odot}$, $M_{\odot}$ and $R_{\odot}$ are the large frequency separation, mass, and radius of the Sun, respectively. P-modes of the same angular degree $l$ are (quasi-)evenly spaced in frequency, and this average spacing is represented by the large frequency separation $\Delta \nu$.

The second scaling relation relates the frequency of maximum oscillation power $\nu_\mathrm{max}$ of p-modes with stellar mass $M$, radius $R$, and effective temperature $T_\mathrm{eff}$:

\begin{equation}\label{eq:astero2}
    \frac{\nu_\mathrm{max}}{\nu_\mathrm{max, \odot}} \simeq \left(\frac{M}{M_{\odot}}\right) \left(\frac{R}{R_{\odot}}\right)^{-2}
    \left(\frac{T_\mathrm{eff}}{T_\mathrm{eff, \odot}}\right)^{-1/2}.
\end{equation} Here $\nu_\mathrm{max}$ is the frequency where the maximum power in p-modes oscillations is observed. Its proportionality to the acoustic cut-off frequency forms the basis for the above scaling relation.

We refer to $M(\bm{\theta}, \nu)$ as the true PSD. We then use the algorithm in \cite{timmer_1995} to randomize the phase of the Fourier transform corresponding to the true PSD that then generates a time-series through an inverse transform. Note that this algorithm generates an evenly-sampled time-series which we use as a simple case study for testing purposes. Similar case studies can be performed for irregularly-sampled time-series, which we explore in Section \ref{subsec:real_modes} by analysing the Kepler-91 time-series. 

\begin{figure}[ht]
    \centering
    \includegraphics[width=\linewidth]{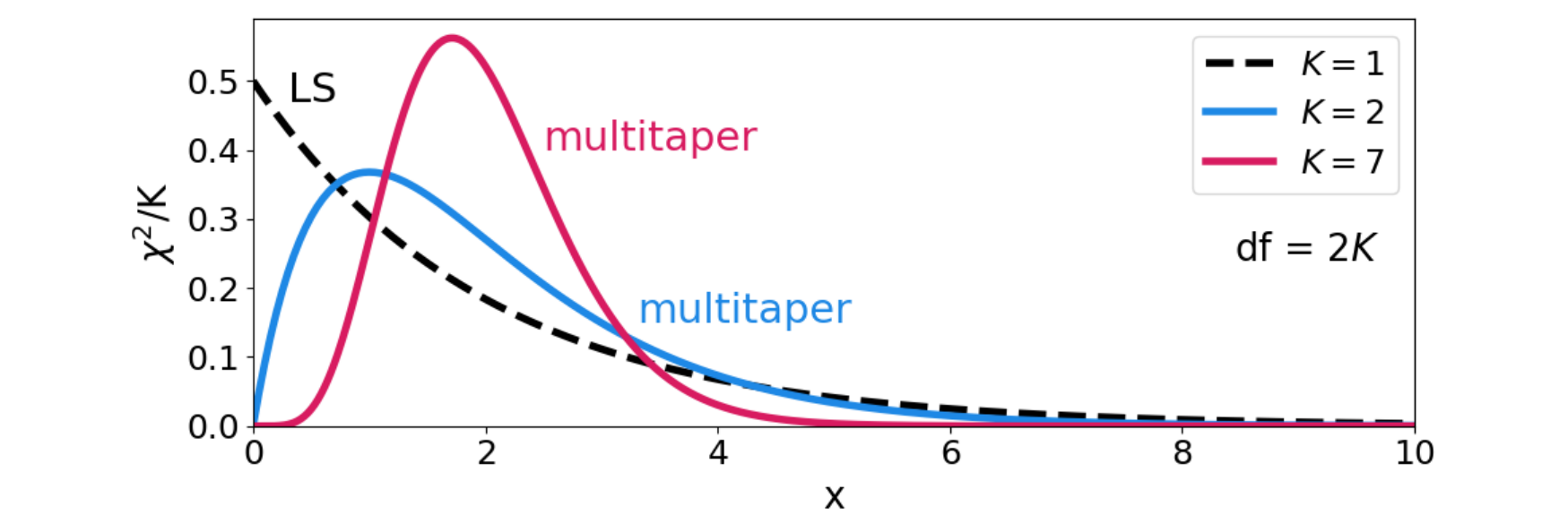}
    \caption{Comparison between distributions of LS and multitaper power spectrum estimates. LS is $\chi^2$ distributed with degrees of freedom $df=2$ as shown by the dashed line in black. \texttt{mtLS} and \texttt{mtNUFFT} are $\chi^2_{2K}$ distributed, where $K$ is the number of tapers. We show the $K=2$ ($\mathrm{df} = 4$) and $K=7$ ($\mathrm{df} = 14$) distributions in blue and dark pink curves. As $K$ (or df) increases, the $\chi^2_{2K}$ approaches a normal distribution with symmetric values around the mean, leading to better noise properties for the \texttt{mtNUFFT} periodogram.}
    \label{fig:chi-square}
\end{figure}

After generating the synthetic light curve, we estimate the true PSD using first the LS and then the \texttt{mtNUFFT} periodograms. We compute two \texttt{mtNUFFT} periodograms, one with bandwidth parameter $NW=3$ and another with $NW=4$. The number of tapers we use follow the $K=2NW-1$ rule. Figure \ref{fig:simulation} compares the \texttt{mtNUFFT} periodograms with LS. We observe the erratic behaviour and spectral leakage of the LS estimate (also shown in Figure 1 of \citealt{anderson_1990}), and the ability of the \texttt{mtNUFFT} periodogram to mitigate these problems. 

The noise in the LS estimate at any given frequency $\hat S^{(\mathrm{LS})}(f)$ is $\chi^2$ distributed with 2 degrees of freedom, whereas that in the \texttt{mtNUFFT} estimate $\hat S^{(\mathrm{mt})}(f)$ is $\chi^2_{2K}$ distributed. As $K$ increases, the $\chi^2_{2K}$ noise distribution approaches a (symmetric) normal, thereby improving upon the large noise values occurring in the $\chi^2_2 \propto e^{-x/2}$ exponential tail. Figure \ref{fig:chi-square} shows these properties of $\chi^2$ distributions. \texttt{mtNUFFT} also reduces out-of-band spectral leakage, and when combined with the F-test in Patil et al. (2024b) improves estimation of (central) frequencies, heights, and widths of the Lorentzians representing p-modes. 

Note that we do not show the low-frequency power excess in Figure \ref{fig:simulation} to focus on mode estimation, but do expect the amplitude of the granulation background (as well as the exponents of the power law fits) to be better estimated using \texttt{mtNUFFT}. A good estimate of the continuum can help deduce granulation and rotational modulation properties \citep{kallinger_2014}, which when combined with mode estimates provide rigorous constraints on stellar models. In the following section, we validate these stellar modeling improvements by applying \texttt{mtNUFFT} to the Kepler-91 light curve. We demonstrate that our method provides precise stellar age estimates for Galactic archaeology studies and improved models of stellar structure and evolution.

\section{Age Estimation}\label{sec:age}

In this paper, we have explored the advantages of multitaper power spectrum analysis for p-mode identification and characterization in red giants and solar-like oscillators. A particularly interesting property of these p-modes is that they are (quasi-)evenly spaced in frequency. Their spacing and location in frequency space has direct connections to fundamental stellar properties like mass, radius, and age. One can demonstrate these connections using the asymptotic theory of stellar oscillations, which is described in detail in \cite{aerts_2010_book} and \cite{chaplin_2011}.

The scaling relations shown in Equations \eqref{eq:astero1} and \eqref{eq:astero2} connect the global asteroseismic properties $\Delta \nu$ and $\nu_\mathrm{max}$ to the mass, radius, and effective temperature of a star \citep[refer to derivation in][]{kjeldsen_1995}. If we add observational constraints such as $T_\mathrm{eff}$ estimates from non-seismic observations, we can solve equations \eqref{eq:astero1} and \eqref{eq:astero2} and estimate stellar mass and radius as follows 

\begin{equation}\label{eq:mass_scaling}
    \frac{M}{M_{\odot}} = \left(\frac{\nu_\mathrm{max}}{\nu_\mathrm{max, \odot}} \right)^3 \left(\frac{\Delta \nu}{\Delta \nu_{\odot}}\right)^{-4} \left(\frac{T_\mathrm{eff}}{T_\mathrm{eff, \odot}}\right)^{3/2}
\end{equation}

\begin{equation}\label{eq:radius_scaling}
    \frac{R}{R_{\odot}} = \left(\frac{\nu_\mathrm{max}}{\nu_\mathrm{max, \odot}} \right) \left(\frac{\Delta \nu}{\Delta \nu_{\odot}}\right)^{-2} \left(\frac{T_\mathrm{eff}}{T_\mathrm{eff, \odot}}\right)^{1/2}.
\end{equation} The mass relation then allows us to estimate precise stellar ages.

If we were to average the large frequency separation between consecutive modes of the same degree $l$, we would get a good estimate of $\Delta \nu$. However, this average $\langle \Delta \nu \rangle$ is sensitive to mode frequency estimates, and any leakage and/or noise in a power spectrum estimate can lead to biased results. The same is true for $\nu_\mathrm{max}$ since it depends on the granulation background and power excess estimates. By reducing spectral leakage and noise (compared to LS), \texttt{mtNUFFT} should improve p-mode characterization, and hence provide accurate estimates of stellar mass, radius, and age through scaling relations. Beyond simple scaling relations, better estimation of mode frequencies and damping rates as well as granulation and/or rotational modulation properties provides fundamental constraints on stellar models and allows us to measure stellar properties with higher accuracy.

In the next section (Section \ref{subsec:real_modes}), we combine the \texttt{mtNUFFT} periodogram estimate of the Kepler-91 light curve with the \texttt{PBjam}\footnote{\url{https://github.com/grd349/PBjam}} Python package to perform peakbagging, i.e., estimate $\Delta \nu$, $\nu_\mathrm{max}$, and independent mode parameters of the red giant. We propagate these estimates to obtain stellar mass, radius, and age, and compare them with those from LS. We demonstrate that peakbagging with \texttt{mtNUFFT} is faster than with LS, thereby allowing large scale asteroseismic analyses using \texttt{PBjam}.

We emphasize that the main result of this paper is the derivation of significant mode frequencies (or mode parameters in general) and deduced $\Delta \nu$ and $\nu_\mathrm{max}$ estimates using \texttt{mtNUFFT}. The additional stellar mass, radius, and age results come from propagating our frequency estimates through existing pipelines that encode the physics of a star. We obtain stellar parameters with better efficiency \textit{solely due to better frequency analysis} rather than the choice of peakbagging algorithm and stellar physics. While we use the \texttt{PBjam} peakbagging algorithm due to its simple, automated, and sound procedure, our method can be extended to any stellar oscillation modeling approach.

\subsection{Kepler-91 Red Giant Time-Series}\label{subsec:real_modes}
We now compare power spectrum estimation using the LS and \texttt{mtNUFFT} periodograms by applying them to same Kepler-91 red giant case study that has been used throughout this paper. We use the following procedure:

\begin{enumerate}
    \item Compute the LS and \texttt{mtNUFFT} periodograms of the stellar light curve
    \item Estimate the radial ($l=0$) and quadropole ($l=2$) mode parameters of the star by analyzing the two power spectrum estimates using the \texttt{PBjam} peakbagging package \label{item:pbjam}
    \item Estimate the stellar mass and radius using the above two sets of mode parameter estimates and scaling relations; then infer stellar age \label{item:pbjam_stellar_infer}
    \item Compare the efficiency, accuracy, and precision of mode and stellar parameter inference (steps \ref{item:pbjam} and \ref{item:pbjam_stellar_infer}) using the two power spectrum estimates
\end{enumerate}

The above procedure directly applies \texttt{PBjam} to both the LS and \texttt{mtNUFFT} power spectrum estimates. While this seems straightforward, there are several statistical assumptions involved that we need to address. We can understand these assumptions by examining the steps involved in \texttt{PBjam} analysis. At its core, \texttt{PBjam} uses a Bayesian approach to fit a solar-like asteroseismic model to the power spectrum estimate of a light curve. It obtains the posterior distribution given the likelihood and the prior distribution
\begin{equation}
    P(\bm{\theta} | D) = P(D |\bm{\theta}) * P(\bm{\theta}) 
\end{equation} 
where $\bm{\theta}$ represents the set of parameters of the asteroseismic model (e.g., Equation \ref{eq:astero_model}), and $D$ is the data that includes the SNR power spectrum estimate. The lightkurve\footnote{\url{https://github.com/lightkurve/lightkurve}} package \citep{lightkurve_2018} generates this SNR estimate by dividing the periodogram power by an estimate of the background (a flattened periodogram). For more details on the preprocessing, refer to \cite{lightkurve_2018}. 

\texttt{PBjam} automates Bayesian fitting of oscillations in three major steps: \texttt{KDE}, \texttt{Asy\_peakbag}, and \texttt{Peakbag}. The first step computes a kernel density estimate (KDE) of the prior $P(\bm{\theta})$ using $\bm{\theta}$ fits to a sample of 13,288 Kepler stars. Figure 3 in \citealt{nielson_2021} shows the distribution of $\bm{\theta}$ for this sample. The prior essentially constrains $T_\mathrm{eff}$, $\Delta \nu$, $\nu_\mathrm{max}$, and a few other parameters to incorporate information on how mode frequencies of stars change with different evolutionary stages.

In addition to generating a prior function, the \texttt{KDE} step provides an initial guess for the parameters $\bm{\theta}$ that are later fit in the asymptotic peakbagging step (referred to as \texttt{Asy\_peakbag}). This initial guess is an estimate of the most probable $\bm{\theta}$ for starting peakbagging. This starting point is estimated by using the KDE estimate of the prior and some input parameters that we need to separately provide to \texttt{PBjam}. The inputs we provide are $T_\mathrm{eff} = 4643.4 \pm 67.3$ \citep[APOKASC-2;][]{pinsonneault_2018}, $\Delta \nu = 9.48 \pm 0.88 \, \mu\mathrm{Hz}$, and $\nu_\mathrm{max} = 109.4 \pm 6.1 \, \mu\mathrm{Hz}$ \citep{lillo_2014}. Note that these values are directly inputted as approximate means and uncertainties of $\bm{\theta}$ and are distinct from those used to estimate the prior. 

The \texttt{KDE} step remains the same for both \texttt{mtNUFFT} and the standard LS power spectrum estimates, so we refer the reader to \citealt{nielson_2021} for more details. On the other hand, we need to evaluate the likelihood for the two power spectrum estimates to be used in the \texttt{Asy\_peakbag} and \texttt{Peakbag} steps. We discuss this likelihood as follows.

\begin{figure*}[ht]
    \centering
    \includegraphics[width=\linewidth]{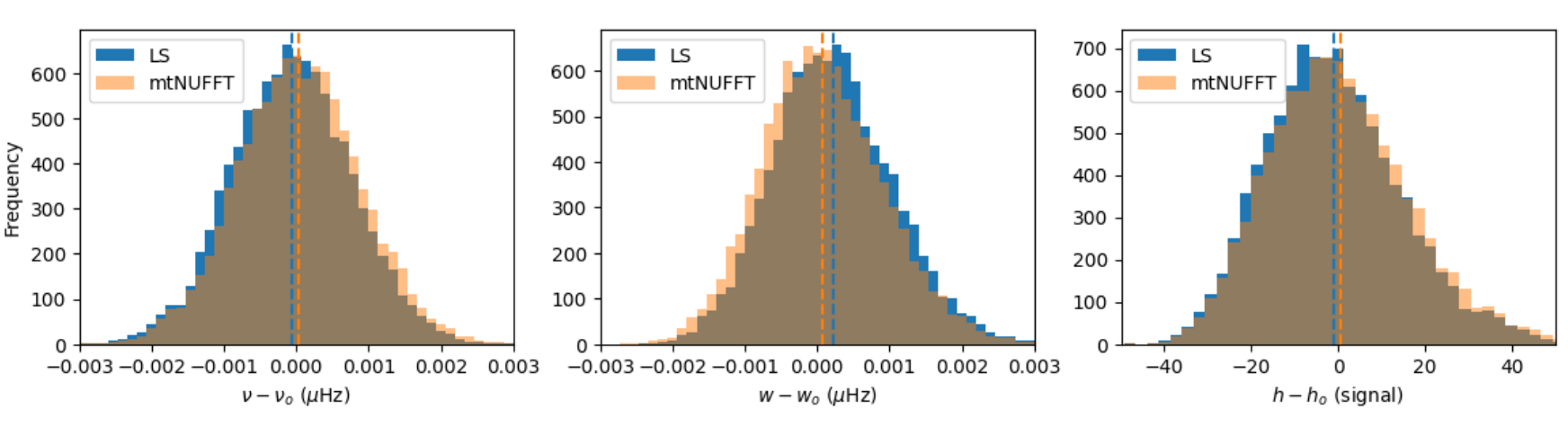}
    \caption{Distributions of best-fit parameters for realizations from a simulated p-mode with frequency $\nu_0$, width $w_0$ and height $h_0$. The left panel shows the deviation of the frequency estimates $\nu$ from the true value $\nu_0$, whereas the middle and right panels show this deviation for the width and height estimates respectively. The blue and orange distributions are obtained by fitting the LS and \texttt{mtNUFFT} periodograms respectively, and their mean values are shown using dotted blue and orange lines. The standard deviations of the two frequency distributions in the left panel are on the order of $8 \times 10^{-4}$, whereas the Libbrecht formula for the precision limit of infinite SNR p-modes is $4 \times 10^{-4}$.}
    \label{fig:single_mode}
\end{figure*}

\begin{enumerate}
    \item \texttt{Asy\_peakbag}: Given the prior $P(\bm{\theta})$ and the starting point for peakbagging (estimated in the \texttt{KDE} step), \texttt{Asy\_peakbag} performs a fit to the asymptotic relation of radial and quadrupole modes (refer to \citealt{chaplin_2011}) by estimating the posterior probability as
    \begin{equation}
        \ln P(\bm{\theta} | D) = \ln \mathcal{L}(\bm{\theta}) + \ln P(\bm{\theta})
    \end{equation} where the log-likelihood is given by
    \begin{equation}
        \ln \mathcal{L}(\bm{\theta}) = \ln \mathcal{L}_{\hat{S}}(\bm{\theta}) + \ln \mathcal{L}_O(\bm{\theta}).
    \end{equation} 
    Here $\ln \mathcal{L}_{\hat{S}}(\bm{\theta})$ is the likelihood of model $M(\bm{\theta}, \nu)$ given SNR power spectrum estimate $\hat{S}_j$ at $j=\{1, \dotsc, J\}$ frequency bins (refer to \citealt{nielson_2021} for information on $\mathcal{L}_O$). 
    For LS power spectrum estimates $\hat S^{(\mathrm{LS})}_j$ that are $\chi^2_2$ distributed\footnote{or Gamma distributed with $\alpha=1$ and $\beta = 1/M(\bm{\theta}, \nu)$} about the expectation $M(\bm{\theta}, \nu_j)$ and are statistically independent across bins $j$, the likelihood is \citep{woodard_1984, duvall_1986, anderson_1990}
    \begin{equation}\label{eq:pbjam_asy_lk}
        \ln \mathcal{L}_{\hat{S}^{(\mathrm{LS})}}(\bm{\theta}) = - \sum\limits_{j=1}^{J} \left( \ln M(\bm{\theta}, \nu_j)  + \frac{\hat{S}^{(\mathrm{LS})}_j}{M(\bm{\theta}, \nu_j)} \right)
    \end{equation}
     This likelihood does not directly apply to \texttt{mtNUFFT} estimates $\hat S^{(\mathrm{mt})}$ as they are $\chi^2_{2K}$ distributed about $M(\bm{\theta}, \nu)$ (refer to Section \ref{subsec:simulation_modes}) and are correlated with neighbouring frequency bins in the range $\nu_j \pm W$. Multitaper spectrum estimates within $2W$ of each other are known to be correlated due to the combined effect of tapers \citep{thomson_2014b}. Beyond this $2W$ frequency range the spectrum estimates are approximately uncorrelated. We perform some tests to confirm that this correlation also applies to \texttt{mtNUFFT}, the results of which are described in Appendix \ref{sec:correlation}.
     
     \cite{anderson_1990} show that the likelihood of a power spectrum estimate that follows a $\chi^2_{2K}$ distribution and is uncorrelated across frequency bins is $K \ln \mathcal{L}_{\hat{S}^{(\mathrm{LS})}}(\bm{\theta})$. Therefore, if the $\chi^2_{2K}$ distributed \texttt{mtNUFFT} estimates were independent across frequency, we would have estimated $\hat{\bm{\theta}}^{(\mathrm{mt})}$ by maximizing the same likelihood as that for LS in Equation \eqref{eq:pbjam_asy_lk}. The uncertainties (or errors) on $\hat{\bm{\theta}}^{(\mathrm{mt})}$ would have then been $\delta \hat{\bm{\theta}}^{(\mathrm{LS})}/\sqrt{K}$. In reality, one needs to correct for the frequency correlations in \texttt{mtNUFFT} estimates by multiplying the uncertainties back up by $\sqrt{K}$. This correction is a good approximation since the correlations introduced by $K$ tapers are close to those given by a boxcar of $K$ bins (refer to the Appendix \ref{sec:correlation}). 
     
     Overall, the result is that the likelihood for \texttt{mtNUFFT} is approximately the same as that for LS, i.e.,
    \begin{align}\label{eq:pbjam_mt_uncer}
        \ln \mathcal{L}_{\hat{S}^{(\mathrm{mt})}}(\bm{\theta}) &= \ln \mathcal{L}_{\hat{S}^{(\mathrm{LS})}}(\bm{\theta})\\
        \Rightarrow \hat{\bm{\theta}}^{(\mathrm{mt})} = \hat{\bm{\theta}}^{(\mathrm{LS})} \;  &\mathrm{and} \; \, \delta \hat{\bm{\theta}}^{(\mathrm{mt})} = \delta \hat{\bm{\theta}}^{(\mathrm{LS})}.
    \end{align} Thus, this step in \texttt{PBjam} does not change for \texttt{mtNUFFT}.
    \label{item:asy_peakbag}
    \item \texttt{Peakbag}: This final step independently fits two Lorentzians to a pair of radial and quadrupole ($l=0, 2$) modes in the power spectrum estimate rather than using the asymptotic relation for the model. It refines the fit by searching for mode parameters centered around the fits obtained in Step \ref{item:asy_peakbag}. The likelihood of this refined model given the $\chi^2_2$ distributed LS estimate stays the same as in Equation \eqref{eq:pbjam_asy_lk}. Thus, this step also does not change for \texttt{mtNUFFT}. \label{item:peakbag}
\end{enumerate}

\begin{figure*}[t]
    \centering
    \includegraphics[width=\linewidth]{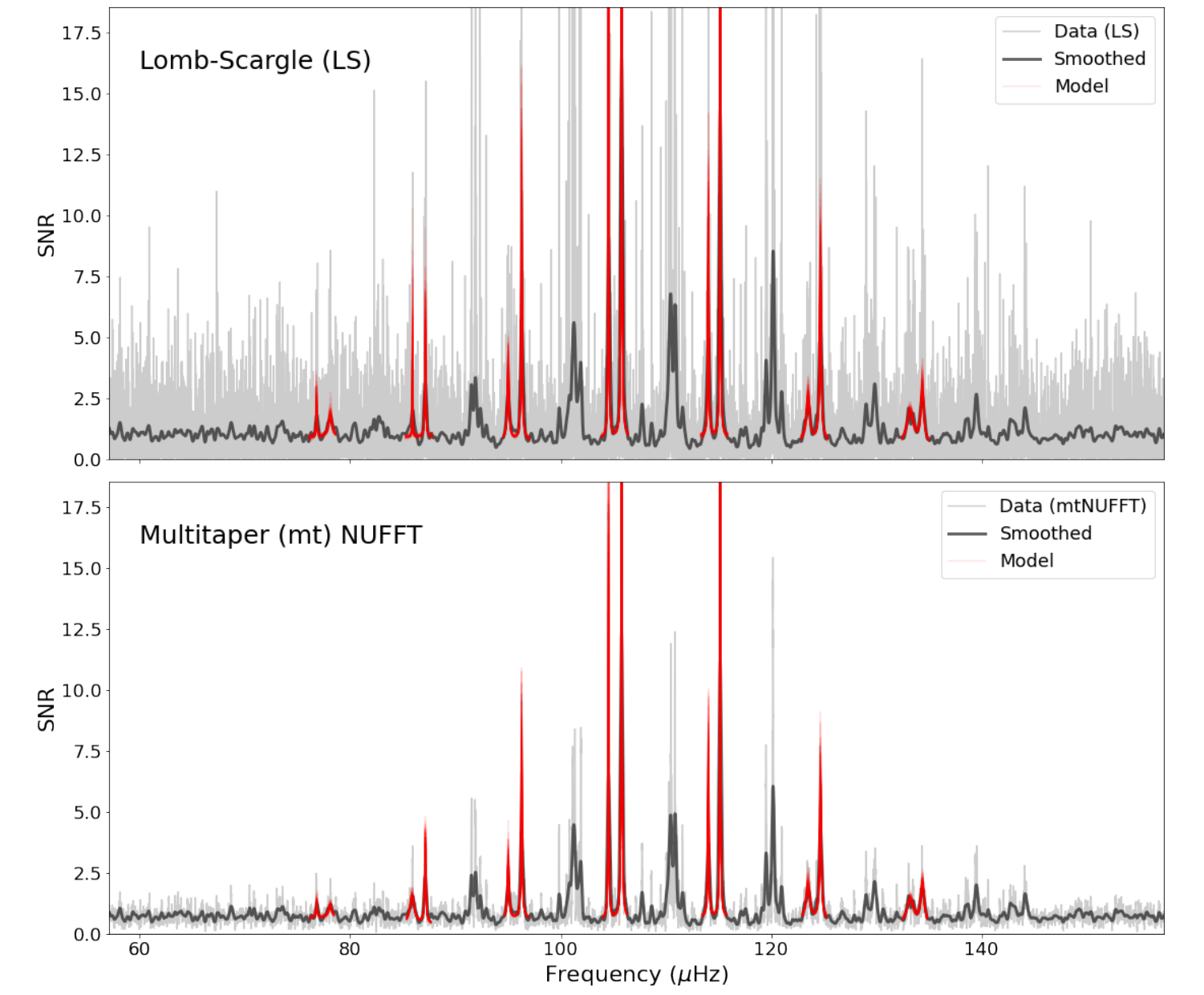}
    \caption{\texttt{PBjam} peakbagging fit for the Kepler-91 time-series using the LS (top panel) and \texttt{mtNUFFT} (bottom panel) periodograms. Both panels show a SNR power spectrum estimate (data) in grey along with its smoothed version using a 1D Gaussian filter kernel in black. The panels also show the model fits (red) to the radial $l=0$ and quadrupole $l=2$ modes obtained in Step \ref{item:peakbag} of \texttt{PBjam}. It is evident that the variance of the \texttt{mtNUFFT} SNR power spectrum estimate is much smaller than that of the LS periodogram, which leads to more efficient peakbagging than LS.}
    \label{fig:pbjam_peakbag}
\end{figure*}

\begin{deluxetable*}{crrrrrrr}
    \tablecaption{\label{tab:mode_parameters}Comparison between \texttt{PBjam} mean and standard deviation estimates of mode parameters using LS and \texttt{mtNUFFT} periodograms. Note that the mode frequency estimates listed here under LS and \texttt{mtNUFFT} have similar mean and standard deviations as those in \citet{lillo_2014}.}
    \tablehead{
    & \multicolumn{3}{c}{mtNUFFT} 
    & \multicolumn{3}{c}{LS}\\
    \hline
    \multicolumn{1}{c}{Degree}
    & \multicolumn{1}{c}{Frequency ($\mu$Hz)}
    & \multicolumn{1}{c}{Width ($\mu$Hz)} 
    & \multicolumn{1}{c}{Height (SNR)} 
    & \multicolumn{1}{c}{Frequency ($\mu$Hz)} 
    & \multicolumn{1}{c}{Width ($\mu$Hz)} 
    & \multicolumn{1}{c}{Height (SNR)}}
    \startdata
     $l=0$ & 78.049 $\pm$ 0.186 & 0.283 $\pm$ 0.275 & 0.3 $\pm$ 0.123& 78.103 $\pm$ 0.093 & 0.386 $\pm$ 0.269 & 0.951 $\pm$ 0.318 \\
           & 87.141 $\pm$ 0.020 & 0.137 $\pm$	0.037 & 2.545 $\pm$ 0.733 & 87.153 $\pm$ 0.019 & 0.119 $\pm$ 0.033 & 4.531 $\pm$ 1.373\\
           & 96.294 $\pm$ 0.017 & 0.132 $\pm$ 0.035 & 6.88 $\pm$ 1.941 & 96.296 $\pm$ 0.016 & 0.128 $\pm$ 0.031 & 9.813 $\pm$ 2.858\\
           & 105.793	$\pm$ 0.010 & 0.103 $\pm$ 0.020 & 20.383 $\pm$ 5.124 & 105.791 $\pm$ 0.010 & 0.097 $\pm$ 0.019 & 27.343 $\pm$ 7.155\\
           & 115.157	$\pm$ 0.009 & 0.088 $\pm$ 0.018 & 19.984 $\pm$ 5.301 & 115.154 $\pm$ 0.010 & 0.089 $\pm$ 0.020 & 26.503 $\pm$ 7.596\\
           & 124.683	$\pm$ 0.020 & 0.169 $\pm$ 0.045 & 6.092 $\pm$ 1.694 & 124.681 $\pm$ 0.019 & 0.156 $\pm$ 0.042 & 8.776 $\pm$ 2.591\\
           & 134.354	$\pm$ 0.047 & 0.255 $\pm$ 0.128 & 1.499 $\pm$ 0.464 & 134.349 $\pm$ 0.042 & 0.213 $\pm$ 0.102 & 2.557 $\pm$ 0.831\\
    \hline
    $l=2$  & 76.779 $\pm$ 0.192 & 0.170 $\pm$ 0.198 & 0.204 $\pm$ 0.091 & 76.811 $\pm$ 0.059 & 0.158 $\pm$ 0.105 & 0.802 $\pm$ 0.305\\
           & 85.92 $\pm$	0.042 & 0.250 $\pm$ 0.143 & 1.364 $\pm$ 0.475 & 85.924 $\pm$ 0.027 & 0.130  $\pm$ 0.091 & 2.718 $\pm$ 1.043\\
           & 95.019 $\pm$ 0.028 & 0.146 $\pm$ 0.059 & 3.295 $\pm$ 1.091 & 95.014 $\pm$ 0.026 & 0.144 $\pm$ 0.052 & 4.892 $\pm$ 1.640\\
           & 104.549 $\pm$ 0.010 & 0.087 $\pm$ 0.020 & 12.186 $\pm$ 3.564 & 104.55 $\pm$ 0.011 & 0.086 $\pm$ 0.020 & 16.64 $\pm$ 4.654\\
           & 114.031 $\pm$ 0.017 & 0.115 $\pm$ 0.035 & 7.375 $\pm$ 2.384 & 114.032 $\pm$ 0.017 & 0.122 $\pm$ 0.035 & 9.724 $\pm$ 3.031\\
           & 123.51 $\pm$ 0.050 & 0.225 $\pm$ 0.121 & 2.51 $\pm$ 0.860 & 123.507 $\pm$ 0.048 & 0.192 $\pm$ 0.105 & 3.733 $\pm$ 1.443\\
           & 133.124 $\pm$ 0.108 & 0.275 $\pm$ 0.227 & 1.034 $\pm$ 0.365 & 133.114 $\pm$ 0.105 & 0.241 $\pm$ 0.223 & 1.794 $\pm$ 0.634\\
    \enddata
\end{deluxetable*}

To check that the likelihood we have derived is correct, we simulate an evenly-sampled realization of a light curve containing a single p-mode oscillation with a linewidth of 1 $\mu$Hz and no background noise. This linewidth is significantly wider than the Rayleigh resolution that sets the intrinsic bin-width in the frequency domain. We then compute the power spectrum of this idealized light curve using both the classical and multitaper periodograms, and fit each spectrum using maximum likelihood estimation (assuming $\chi^2_{2}$ and $\chi^2_{2K}$ distributions for classical and multitaper estimates, respectively). By repeating this experiment a large number of times, on new realizations, we obtain a distribution of best-fitting mode parameters for the LS and the multitaper cases. 

The mode parameter distributions are shown in Figure \ref{fig:single_mode}. We find that the standard deviations of the LS and multitaper distributions are Gaussian and almost identical. We find similar results upon fitting LS and \texttt{mtNUFFT} periodograms of unevenly-sampled p-mode realizations. This suggests that the \texttt{mtNUFFT} likelihood should be the same as that for LS, as predicted in Equation \eqref{eq:pbjam_mt_uncer}. Additionally, the two sets of estimates have almost identical distributions, suggesting that they both do well in the evenly-sampled case of an infinite SNR p-mode. Note that the obtained standard deviations of the best-fit distributions are larger than the underlying precision limit of individual frequencies given by \citet{libbrecht_1992} (see also \citealt{chaplin_2002}).

We now compare \texttt{PBjam} asteroseismic inference of the Kepler-91 light curve using LS and \texttt{mtNUFFT} periodograms. Figure \ref{fig:pbjam_peakbag} shows the peakbagging fit for both periodograms. Smoothing the \texttt{mtNUFFT} SNR power spectrum estimate using a 1D Gaussian filter kernel results in an estimate similar to itself, as the variance of this estimate is small. Note that the kernel used here has a standard deviation $\sigma = 1/\Delta f$, where $\Delta f$ is the frequency sampling of the power spectrum estimate. This is not the case for LS, where smoothing using the same Gaussian filter kernel results in a significant variance reduction. Thus, it is more computationally efficient to perform peakbagging with \texttt{mtNUFFT} rather than LS. We compare the wall-clock time taken by the final peakbagging step \ref{item:peakbag} for the two periodograms, and find that \texttt{mtNUFFT} provides \textit{a factor three speed-up}.

Furthermore, smoothing or averaging the LS periodogram to reduce its variance is not the same as computing a multitaper power spectrum estimate. This is because the smoothed LS estimate averages over signal and leakage leading to false mode detections and inaccurate frequency estimates. Thus, in addition to efficiency, we test the accuracy and precision of estimation. 

Table \ref{tab:mode_parameters} compares the \texttt{PBjam} $l=0, 2$ mode frequency estimates using LS and \texttt{mtNUFFT}. We see that the two sets of \texttt{PBjam} estimates are consistent with each other, and that the $1 \sigma$ uncertainties on the \texttt{mtNUFFT} estimates are similar to those for LS. We also see that there are subtle differences between the LS and \texttt{mtNUFFT} (mean) mode frequency estimates, which could be because of the variance (noise) reduction provided by \texttt{mtNUFFT}. These subtle differences might also be due to the \texttt{mtNUFFT} estimates being slightly less biased than those of LS (see Figure~\ref{fig:single_mode}).

Along with frequency, \texttt{PBjam} infers peak width and height of p-modes, which are also presented in Table \ref{tab:mode_parameters}. Given a multitaper power spectrum estimate, these inferred mode parameters should be dependent on the bandwidth $W$ that defines the power spectral window. Usually, one uses a reasonably small $W$ to reduce the effect of such dependency. For example, if a mode is split between several bandwidths, then we can obtain a good estimate of the peak width, especially due to the reduced bias (out-of-band leakage) of \texttt{mtNUFFT}. Otherwise, the precision is limited by the local bias within $f \pm W$, which leads to rectangular-shaped peaks in the multitaper estimate. Similarly, improved height estimation follows from the reduced bias of \texttt{mtNUFFT}, but can also be affected by local bias (see Figure \ref{fig:simulation}). 

One can improve mode width and/or height precision by using approaches that tackle local bias \citep[][]{riedel_1995, prieto_2007}. Using helioseismic time-series, \cite{korzennik_2005} present a method to optimally choose the number of tapers $K$ and correct for any bias in multitaper line width estimates \citep[see also][]{komm_1999}. In the future, we would like to combine this method with \texttt{mtNUFFT}.

\texttt{PBjam} also provides estimates of average seismic parameters. We find that these estimates obtained using the LS and \texttt{mtNUFFT} periodograms (shown in Table \ref{tab:mode_parameters}) are consistent and have similar precision, even more so than the individual mode frequency estimates. The smaller differences are because these properties are estimated by averaging over several modes. Thus, the inferred estimates of bulk stellar properties like mass and radius are similar for \texttt{mtNUFFT} and LS when using scaling relations. Note that both our \texttt{PBjam} estimates are also consistent with those from the APOKASC-2 sample: $\nu_\mathrm{max} = 109.445 \pm 0.985 \, \mu\mathrm{Hz}$ and $\Delta \nu = 9.437 \pm 0.038 \, \mu\mathrm{Hz}$ \citep{pinsonneault_2018}. This sample combines Kepler asteroseismic time-series with the APOGEE spectroscopic sample to estimate average asteroseismic parameters. In particular, it calibrates and combines estimates obtained from several different pipelines.

\begin{deluxetable*}{rrrrrrrr}
    \tablecaption{Comparison between \texttt{PBjam} mean and standard deviation estimates of average seismic and stellar parameters using LS and \texttt{mtNUFFT} periodograms. These estimates are also compared with the \texttt{APOKASC-2} uncorrected (u) and corrected (c) scaling relation estimates. \label{tab:average_mode}}

    \tablehead{
    & \colhead{$\nu_\mathrm{max}$ ($\mu\mathrm{Hz}$)} 
    & \colhead{$\Delta \nu$ ($\mu\mathrm{Hz}$)}
    & \colhead{Mass ($\mathrm{M}_\odot$)} 
    & \colhead{Radius ($\mathrm{R}_\odot$)}
    & \colhead{Log Age ($\mathrm{Myr}$)}
    & \colhead{\% Age}}
    \startdata
    \texttt{mtNUFFT} & 110.090 $\pm$ 0.720 & 9.432 $\pm$ 0.007 & 1.375 $\pm$ 0.040 & 6.557 $\pm$ 0.064 & 3.599 $\pm$ 0.053 & 13.07\\
    LS               & 109.826 $\pm$ 0.711 & 9.429 $\pm$ 0.008 & 1.366 $\pm$ 0.040 & 6.545 $\pm$ 0.064 & 3.607 $\pm$ 0.056 & 13.70\\
    (c)\texttt{APOKASC-2} & 109.445 $\pm$ 0.985 & 9.437 $\pm$ 0.038 & 1.219 $\pm$ 0.046 & 6.181 $\pm$ 0.087 & 3.832 $\pm$ 0.061 & 15.08\\
    (u)\texttt{APOKASC-2} & & & 1.347 $\pm$ 0.051 & 6.511 $\pm$ 0.091 & 3.630 $\pm$ 0.070 & 17.53
    \enddata
\end{deluxetable*}

The standard deviations on the \texttt{PBjam} average asteroseismic estimates are smaller than the uncertainties on the APOKASC-2 estimates, thereby illustrating that \texttt{PBjam} peakbagging provides precise estimates. As shown in Table \ref{tab:average_mode}, the stellar mass and radius estimates derived using scaling relations with \texttt{PBjam} have smaller uncertainties as compared to APOKASC-2. We also see that the mass and radius uncertainty reduction is smaller compared to that of $\Delta \nu$ and $\nu_\mathrm{max}$. We can understand this by propagating $\Delta \nu$ and $\nu_\mathrm{max}$ uncertainties into the mass and radius scaling relations \eqref{eq:mass_scaling} and \eqref{eq:radius_scaling}. The following mass uncertainty formula is derived using error propagation through partial derivatives with the assumption that the uncertainties on $\Delta \nu$, $\nu_\mathrm{max}$ and $T_\mathrm{eff}$ are small

\begin{align*}   
\sigma_{M/M_\odot}^2 =  \left(\frac{M}{M_\odot}\right)^2 &\left[ 9\left( \frac{\sigma_{\nu_\mathrm{max}}}{\nu_\mathrm{max}}\right)^2 + 16\left( \frac{\sigma_{\Delta \nu}}{\Delta \nu}\right)^2 \right. \\ &\left. + 2.25\left( \frac{\sigma_{T_\mathrm{eff}}}{T_\mathrm{eff}} \right)^2 \right]. \numberthis \label{eq:mass_uncert}
\end{align*}

Thus, the uncertainty on stellar mass is dominated by the fractional uncertainties of $\Delta \nu$ and $\nu_\mathrm{max}$ with factors 16 and 9 respectively in Equation \eqref{eq:mass_uncert}. However, for our case study of Kepler-91, these uncertainties are very small, on the order of $0.08$ and $0.6$\%, respectively. In contrast, the $T_\mathrm{eff}$ fractional uncertainty is ${\approx}1.45\%$, which contributes more to the total mass error despite its $2.25$ factor in Equation \eqref{eq:mass_uncert}. The same is true for stellar radius uncertainty. Instead of directly using the formula in Equation \eqref{eq:mass_uncert} to list mass uncertainties in Table \ref{tab:average_mode}, we estimate these uncertainties by drawing $\Delta \nu$, $\nu_\mathrm{max}$, and $T_\mathrm{eff}$ samples from normal distributions with means and standard deviations given in Table \ref{tab:average_mode} (and $T_\mathrm{eff} = 4643.4 \pm 67.3$ in APOKASC-2) and applying uncorrected scaling relations. We then confirm that these uncertainties are consistent with Equation \eqref{eq:mass_uncert}. We repeat this procedure for stellar radius estimates.

Finally, we propagate the stellar mass uncertainty to age. We use the \texttt{scipy} piecewise linear interpolation on the APOKASC-2 sample to estimate their mapping from (mass, $[\mathrm{Fe/H}]) \rightarrow$ age. This empirically approximates the stellar age function $f(\mathrm{mass}, [\mathrm{Fe/H}]))$ using the stellar models computed by APOKASC-2. We then compute the implied age of Kepler-91 using our mass estimates and $[\mathrm{Fe/H}]$ estimates from APOKASC-2. The uncertainties are computed in the same way we compute age and radius uncertainties, i.e., by sampling normal distributions with means and standard deviations given by corresponding estimates of mass and $[\mathrm{Fe/H}]$. We compare our \texttt{PBjam} age estimates with the APOKASC-2 age estimates using uncorrected scaling relations and those with corrections applied (refer to \citealt{pinsonneault_2018} for more details). The age estimates obtained using \texttt{PBjam} with LS and \texttt{mtNUFFT} are shown in Table \ref{tab:average_mode}. These age estimates are more precise than those from APOKASC-2, whose age precision is $17.53$\% (and $15.08$\% with applied corrections).

Note that the uncorrected scaling relations \eqref{eq:mass_scaling} and \eqref{eq:radius_scaling} assume that we can scale all stars with solar-like oscillations to the Sun, an approximation that does not entirely hold for the evolved stars. For example, the $l=1$ modes in red giants have mixed p and g-mode characteristics. These mixed mode frequencies and widths are hard to estimate and are thus not yet included in \texttt{PBjam}. We expect that \texttt{mtNUFFT} will provide more accurate stellar property estimates if stellar models are constrained using independent frequency estimates, including the $l=1$ modes. In the future, we would like to use these independent frequencies instead of scaling relations to study the impact of multi-tapering on age estimation. We would also like to perform more detailed analyses for testing potential improvements in bias due to the use of \texttt{mtNUFFT}. For example, we could test whether tapers help return a higher fraction of reliable fits when the modes have low SNR.

\section{Discussion \& Conclusion}\label{sec:discussion}
The main takeaways from this paper are:

\begin{enumerate}
    \item \texttt{mtNUFFT} estimates power spectra of time-series data with \textit{reduced bias and variance} than those estimated using the LS periodogram.
    \item In combination with the \texttt{PBjam} algorithm, \texttt{mtNUFFT} provides a factor of three speed-up in modelling solar-like oscillations as compared to the LS. This allows efficient estimation of precise asteroseismic ages that can be scaled to a large number of stars. Note that the \textit{efficiency} improvement is a consequence of \textit{variance reduction}.
    \item The \textit{bias suppression} provided by \texttt{mtNUFFT} should improve the \textit{accuracy} of frequency estimates of solar-like oscillations. Our companion paper, Patil et al. (2024b), extends \texttt{mtNUFFT} to the multitaper F-test to also improve the \textit{precision} of oscillation frequencies.
\end{enumerate}

Thus, \texttt{mtNUFFT} is a novel frequency analysis method that can provide major improvements in asteroseismology, stellar and Galactic archaeology studies, and time-domain astronomy in general. We discuss these below.
 
\textbf{\textit{Prospects for Asteroseismology:}} Our multitaper power spectrum estimation method, \texttt{mtNUFFT}, in conjunction with a peakbagging algorithm, can precisely find the center frequencies of Lorentzians that represent p-modes. We can use these individual mode estimates to model stars as well as test stellar structure and evolution theory. Additionally, this would improve age estimation accuracy of solar-type and red-giant stars beyond that of simple mass scaling relations, which has promising implications for understanding Galaxy formation and evolution \citep{chaplin_2013}. 

In Section \ref{subsec:real_modes}, we only dealt with radial and quadrupole p-modes in Kepler-91. If we were to improve the frequency and width precision of $l=1$ mixed modes in red giants using \texttt{mtNUFFT}, we would be able to probe stellar cores in detail \citep{bedding_2011, mosser_2017, hekker_2017}. Improving width precision of mixed modes, and p-modes in general, requires methods that tackle local bias in \texttt{mtNUFFT} estimates, e.g., the minimum bias and sinusoidal tapers in \cite{riedel_1995} or the quadratic multitaper approach \citep{prieto_2007}. We would like to extend these methods to \texttt{mtNUFFT} in the future.

We point the reader to our companion paper (Patil et al. 2024b), where we focus on harmonic analysis for the detection and precise frequency estimation of strictly periodic signals in time-series data. Essentially, we extend the Thomson F-test \citep{thomson_1982} to our \texttt{mtNUFFT} periodogram and show that it automatically detects the Kepler-91b exoplanet transit harmonics \citep{lillo_2014} in the Kepler-91 time-series. The Kepler-91 transit example demonstrates that the F-test could be used to distinguish between p, g, and undamped modes in stars as well. In Patil et al. (2024b), we also demonstrate the technique of dividing a time series into chunks and examining each chunk separately to understand how periodic signals change over time. In particular, we apply the \texttt{mtNUFFT}/F-test combination to individual chunks of the Kepler-91 light curve to investigate the periodic versus transient nature of oscillatory modes.

In the future, we plan to extend \texttt{mtNUFFT} to estimate masses and ages of red giants in old open clusters to better understand mass-loss along the red giant branch \citep{miglio_2012, handberg_2017} as well as to investigate the overall improvement in stellar age precision provided by our method \citep{shu_1987, lada_2003}. We also aim to develop a precise stellar age catalog by applying our method to a large number of stars in the Kepler field, e.g., the APOKASC-3 sample.

\textbf{\textit{Comparison with other methods:}} The multitaper approach has similarities with other power spectrum estimators that tackle bias and consistency. One such example is the Welch periodogram \citep{welch_1967} that divides a time-series into overlapping sections, tapers and calculates the periodogram of each section, and then averages these periodograms \citep[refer to applications in][]{dodson-robinson_2022}. The tapering and overlapping of sections ensures bias suppression and independence of periodograms for variance reduction. However, the frequency resolution of this power spectrum estimator is limited by the length of the sections. On the other hand, the multitaper estimate applies orthogonal DPSS tapers to entire time-series, thereby providing a higher frequency resolution than the Welch estimate with the same bias and variance. Additionally, if you fix frequency resolution and bias (or variance), the multitaper estimator has smaller variance (or bias) \citep{bronez_1992}.

Note that the spline interpolation in \cite{springford_2020} could be improved for accuracy while still maintaining its computational gain over the generalized DPSS for irregular sampling \citep{bronez_1988}. \cite{chave_2019} develop an efficient multitaper estimator for time-series with missing data that does not interpolate tapers. We aim to compare our approach with theirs to see how much quasi-regularity and large gaps in time affect the results.

\textbf{\textit{Summary:}}
The LS periodogram is a widely-used power spectrum estimator for unevenly-sampled time-series analysis, particularly in asteroseismology. However, this periodogram suffers from the statistical issues of bias and inconsistency. We improve upon these issues by combining the multitaper statistic for evenly-sampled time-series with the \texttt{NUFFT} periodogram. Here, multitaper refers to windowing of time-series using DPSS tapers that minimize spectral leakage (bias) and variance (inconsistency) in a power spectrum estimate, whereas the \texttt{NUFFT} periodogram is the generalization of the classical periodogram that works well for quasi-regular time sampling such as that of space-based light curves.

Using simulations and the case study of the Kepler-91 red giant, we show that \texttt{mtNUFFT} is more efficient than LS at providing high-precision frequency estimates of solar-like oscillations using \texttt{PBjam} peakbagging. We also expect better characterization of p-modes when using \texttt{mtNUFFT} due to the bias reduction it offers, as suggested by the simulation results\footnote{a more rigorous analysis is required to demonstrate accuracy improvement}. For Kepler-91 in particular, we obtain an age estimate of $3.97 \pm 0.52$ Gyr. While the improvement in precision as compared to the APOKASC-2 (uncorrected) estimate of $4.27 \pm 0.75$ Gyr is due to the use of \texttt{PBjam}, combining \texttt{mtNUFFT} with \texttt{PBjam} increases the efficiency of peakbagging as compared to LS. Thus, our method can provide major improvements to asteroseismology, stellar structure and evolution models, and Galactic archaeology.

Note that the advantages of our method are not limited to the above studies. We present a new and powerful frequency analysis method that generally applies to time-domain astronomy, and envision that it will prove beneficial for upcoming surveys such as the Vera Rubin Observatory Legacy Survey of Space and Time \citep{lsst_2009}. We also encourage the reader to look at our companion paper, Patil et al (2024b), where we present the harmonic F-test for the \texttt{mtNUFFT}. The public Python package, \texttt{tapify} (refer to Appendix \ref{sec:tapify}), aids in the application of our method across astronomy and different fields of science and engineering.

\begin{acknowledgements}
AAP is supported by an LSST-DA Catalyst Fellowship; this publication was thus made possible through the support of Grant 62192 from the John Templeton Foundation to LSST-DA. AAP and this project received support from the Data Sciences Institute at the University of Toronto (UofT). GME acknowledges funding from NSERC through Discovery Grant RGPIN-2020-04554 and from UofT through the Connaught New Researcher Award, both of which supported this research.

The authors thank the anonymous referee for their thorough and thoughtful feedback. They also thank Conny Aerts and Jo Bovy for helping design the project and providing insightful feedback on the manuscript, Bill Chaplin for helping fix issues with likelihood calculations, and Ted Mackereth for providing advice on light curve simulation.

This paper includes data collected by the Kepler mission and obtained from the MAST data archive at the Space Telescope Science Institute (STScI). The data can be found in MAST: \dataset[10.17909/b90b-a558]{http://dx.doi.org/10.17909/b90b-a558}. Funding for the Kepler mission is provided by the NASA Science Mission Directorate. STScI is operated by the Association of Universities for Research in Astronomy, Inc., under NASA contract NAS 5–26555.

\software{\texttt{astropy} \citep{astropy:2013, astropy:2018, astropy:2022}, \texttt{FINUFFT} \citep{barnett_2019}, lightkurve \citep{lightkurve_2018} \texttt{matplotlib} \citep{matplotlib:2007}, \texttt{nfft} \citep{nfft_2017}, \texttt{numpy} \citep{numpy:2020}, \texttt{pbjam} \citep{nielson_2021}, \texttt{scipy} \citep{scipy:2020}.}

\end{acknowledgements}

\appendix

\section{\texttt{tapify}}\label{sec:tapify}
We develop \texttt{tapify} \citep{tapify}, a Python package for multitaper spectral analysis, that is generally applicable to time-domain astronomy. \texttt{tapify} takes inspiration from previously written R packages. These are \texttt{multitaper} \citep{multitaper}, which provides methods for evenly-sampled time-series, and \texttt{mtLS} \citep{springford_2020}, which implements the \texttt{mtLS} periodogram for uneven-sampling. The \texttt{mtLS} package extends the \cite{multitaper} package to uneven-sampling, but does not include the F-test and other features such as adaptive weighting and jackknife variance/confidence intervals. We thus extend the \texttt{mtLS} package to include these features as well as the \texttt{mtNUFFT} periodogram discussed in Section \ref{subsubsec:mtnufft}. We generate all the figures in this paper (except Figure \ref{fig:kepler_time}) using this package.

We also add metrics and algorithms that help choose the multitaper $NW$ and $K$ parameters in the package, which we discuss in Appendix \ref{sec:nw_k_choose}. To aid with testing of asteroseismic analyses, we provide methods to simulate a time-series given a theoretical power spectrum. We simulate time-series of p, g, and coherent quasi-infinite lifetime modes, apply \texttt{mtNUFFT} spectral analysis techniques, and analyze the results. The \texttt{mtNUFFT} periodogram accurately and precisely recovers the true power spectra, and its F-test preferentially picks up coherent and g-modes as opposed to p-modes. We describe the simulation methods in Section \ref{subsec:simulation_modes} and use them to test stellar age estimation using asteroseismic time-series.

To aid with the usage of \texttt{tapify}, we provide the following workable example:

\begin{lstlisting}[language=Python]
from tapify import MultiTaper

# Read the time-series
t = ...
y = ...

# Set the parameters: bandwidth NW and the number of tapers K
NW = 4
K = 2 NW - 1

# Create a MultiTaper object of the given time-series
mt_object = MultiTaper(y, t=t, NW=NW, K=K)

# Compute the multitaper NUFFT periodogram
freq, power, f_statistic = mt_object.periodogram(method='fft', adaptive_weights=True,
                                                 jackknife=True, ftest=True)
\end{lstlisting}

For more details, refer to the GitHub repository of \texttt{tapify} at \url{https://github.com/aaryapatil/tapify} and its documentation at \url{https://tapify.readthedocs.io/}.

\section{Choosing NW and K}\label{sec:nw_k_choose}
The one caveat of multitaper power spectrum estimators is the trade-off between statistical stability and frequency resolution, which one tunes using the time-bandwidth product ($NW$) and the number of tapers ($K$) \citep{thomson_1982, springford_2020}. While obtaining an unbiased estimate of the power spectrum underlying an observed time-series is not possible, we can choose $NW$ and $K$ for multitaper power spectrum estimates to approximately attain the desired statistical properties. In particular, as $NW$ increases, more tapers ($K \approx 2NW$) with large in-band power concentration and minimal spectral leakage out-of-band are available for estimating independent power spectrum estimates $\hat S^{(mt)}_k(f)$, thereby controlling the variance of the averaged $\hat S^{(mt)}(f)$ estimate. However, increasing $NW$ also leads to larger local bias that results in reduced frequency resolution (refer to Figure \ref{fig:pseudo_window_mt}). In short, larger $NW$ preferentially reduces variance and spectral leakage (out-of-band bias) over frequency resolution (local bias). Based on these considerations, one can in general use a larger $W$ as the sample size $N$ increases \citep{haley_2017}.

Since the frequency resolution is sensitive to the choice of bandwidth, one may directly choose $NW$ based on the resolution required for a particular study. For example, if two asteroseismic modes are spaced $2 W_\star$ Hz apart in frequency and we wish to resolve them, we must choose the $W < W_\star$. However, \cite{haley_2017} demonstrate that choosing a very small or large $W$ can have adverse effects on multitaper power spectrum estimates. Therefore, we use methods from the statistics literature for bandwidth or $W$ selection, and aim to strike a balance between desired (theoretical) frequency resolution and statistical stability.

While there is no standard method for choosing the optimum $NW$ and $K$ for any given problem, some studies provide metrics \citep[e.g.,][]{thomson_1982} and algorithms \citep[e.g.,][]{haley_2017} for tuning these parameters based on some assumptions about the process underlying the time series. 

We discuss two approaches below
\begin{enumerate}
    \item  \cite{thomson_1982} introduce the stability estimate $\upsilon(f)$ and variance efficiency $\Xi_K$ that help us choose $NW$ and $K$ for a given problem. The stability estimate is given by
    
    \begin{equation}\label{eq:stab_est}
    \upsilon(f) = 2 \sum_{k=0}^{K-1} \left| d_k(f) \right|^2
    \end{equation} which is approximately the degrees of freedom of $S^{(\mathrm{mt})}(f)$. an indicator of the amount of bias in a power spectrum estimate. If $\bar{\upsilon}(f)$, the frequency-averaged $\upsilon(f)/2K$, is $\ll 1$, then the bias is expected to be too high and this could be due to a very small $W$. Thus, one can estimate $\upsilon(f)$ for a range of $W$, to choose the \textit{best} value. 

    The variance efficiency $\Xi_K$ is a measure of variance and covariance of a power spectrum estimate across frequencies. It is a real number between 0 and 1, 1 being the most efficient, and is defined as
    \begin{equation}\label{eq:var_eff}
    \Xi_K = \frac{1}{N \sum\limits_{n=0}^{N-1} \left[\frac{1}{K} \sum\limits_{k=0}^{K-1} {[ v_{k, n}(N, W)]}^2 \right]^2},
    \end{equation}
    One drawback of this metric is that it relies on the assumption of white noise (the power spectrum of a ``white" noise process is constant, e.g., an independent and identically–distributed process), and does not take bias into consideration. Thus, it should be used in combination with other metrics to compare power spectrum estimates \citep{thomson_1982}. Particularly, we can combine the stability estimate for bias protection along with the variance efficiency to get an overall efficiency measure
    \begin{equation}\label{eq:eff}
    \mathrm{eff} \approx \bar{\upsilon}(f) \, \Xi_K \, N
    \end{equation} which can be used to compare different $W$ and $K$. An optimal value of $\mathrm{eff}$ would be $\approx N$.
    
    Note that the variance efficiency does not depend on a particular time-series; it is calculated using the grid approximations to the DPSS tapers $\mathbf{v}(N, W)$. On the other hand, the stability estimate uses adaptive weights $d_k(f)$ that minimally depend on the observed time-series and resemble $\lambda_k$. Thus, these metrics are more general recommendations rather than strict criteria for $NW$ and $K$ selection. Table II in \cite{thomson_1982} shows how the above metrics change with $NW$ and $K$. Typically, $NW \approx 4$ is a good choice. For instance, $NW$ = 4, $K=5$ has $81.4$\% variance efficiency \citep{thomson_1982} and a generally high stability estimate.
    
    \item \cite{haley_2017} propose a systematic method to obtain the optimal bandwidth for multitaper power spectrum estimation given a time-series. The method minimizes the Mean Squared Error (MSE) of the log power spectrum, which is a combination of its squared (local) bias estimate and the variance estimate. Assuming that the true power spectrum is smooth, one can estimate the local bias using a spline approximation that determines the curvature of the power spectrum. The variance estimate is obtained using the jackknifing technique in \cite{thomson_1991} (refer to Section \ref{subsubsec:mt}). 
    
    Since this method requires smooth spectra without any purely periodic signals, it can only be used for analyzing asteroseismic time-series that are known to have a smooth granulation background and a comb-like p-mode pattern, i.e., those of solar-like oscillators. However, since we cannot rule out the possibility of the presence of sinusoidal signals due to extrinsic features such as exoplanets, a good test is to use $NW$ and $K$ based on the Thomson metrics, estimate the multitaper F-test and ensure that no periodic signal is present. Another potential method is to perform the multitaper F-test on a given time-series and then remove periodic signals from the power spectrum using a prewhitening approach. In the Kepler-91 case, this method does not work well because there are several periodic signals with uncertainties in their frequency estimates. We provide this optimization technique in the \texttt{tapify} package but caution that it should only be used if appropriate.
    
    \item We can also directly use the harmonic F-test (see our companion paper, Patil et al. 2024b) to make informed decisions about bandwidth selection. We can analyze pseudowindows with varying bandwidth and try detecting injected signals using the F-test. If $NW$ is too narrow, the F-test will be unable to detect the signal; if it is too wide, there will be spurious detections. $NW=4$ seems to work well for our case study.
    \end{enumerate}
    
    We expand upon bandwidth selection in Section \ref{subsec:simulation_modes}, where we use $NW=3$ and $NW=4$ to see how the frequency resolution changes and the effect it has on asteroseismic mode detection. Note that we set $K$ to $2NW - 1$ to control out-of-band spectral leakage, but this discrete parameter could be tuned to further minimize leakage.

\section{Jackknife variance estimator for \texttt{mtNUFFT}}\label{sec:jackknife}
\begin{figure}[ht]
    \centering
    \includegraphics[width=0.75\linewidth]{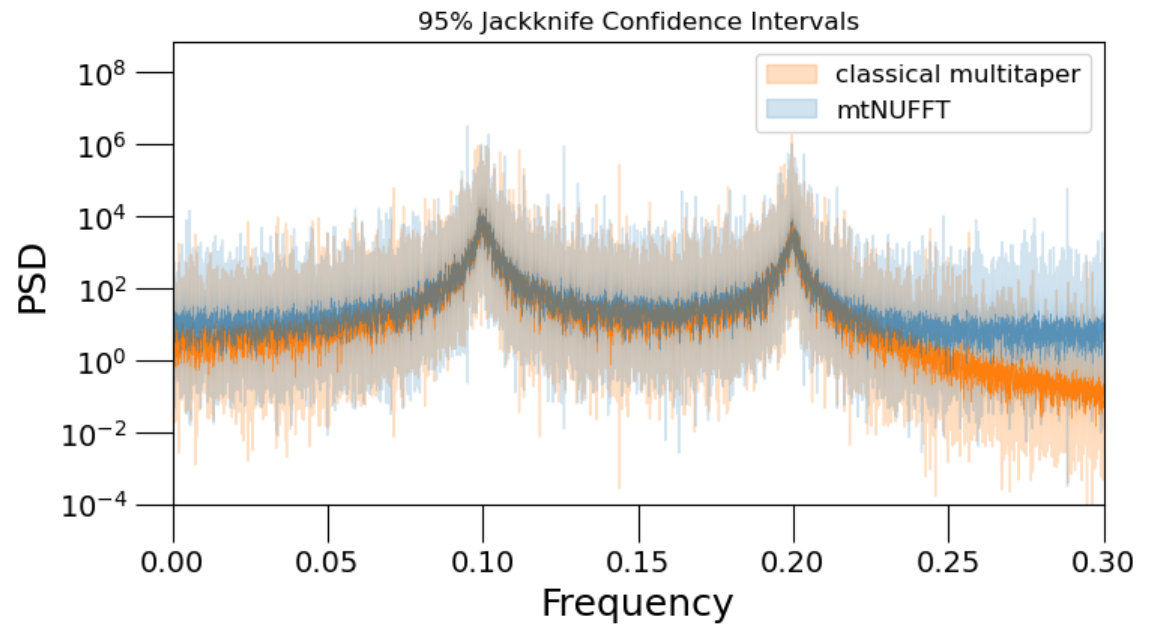}
    \caption{Comparison between 95\% jackknife confidence intervals of the classical multitaper (transparent orange) and \texttt{mtNUFFT} (transparent blue) power spectrum estimates of regular and quasi-regular AR(4) process ($N = 44544$) time-series respectively. The confidence intervals are consistent with each other showing that interpolation of DPSS tapers to Kepler-like time-series introduces minor non-orthogonality, and allows us to compute the jackknife variance. Note that the high frequency power spectrum estimates have larger differences because of spectral leakage due to uneven time sampling.}
    \label{fig:jackknife_compare}
\end{figure}

One can compute the jackknife variance of the classical multitaper power spectrum estimator because the individual eigenspectra (single-tapered estimates) $\hat{S}_k(f)$ are independent due to the orthogonality of DPSS tapers. However, interpolation of DPSS tapers to quasi-regularly sampled time-series makes them slightly non-orthogonal, thereby making the jackknife variance estimator potentially inapplicable to \texttt{mtNUFFT}. Here we show that for Kepler-like time-series the non-orthogonality is minor and therefore jackknife estimates are still valid. Particularly, we conduct the following experiment. First, we simulate an autoregressive process (AR) of order $4$, sample it at timestamps similar to Kepler-91, and estimate its power spectrum using \texttt{mtNUFFT} along with a jackknife confidence interval. Then, we sample the same process on an evenly-spaced grid with $\Delta t$ equal to the mean sampling interval $\overline{\Delta t}$ of Kepler-91, and compute the classical multitaper estimate and its confidence interval. We find that the confidence intervals for the classical multitaper and \texttt{mtNUFFT} estimates are very similar. Figure \ref{fig:jackknife_compare} shows such a comparison.

\section{Correlation of \texttt{mtNUFFT} estimates across frequency}\label{sec:correlation}

\begin{figure*}[htbp]
    \centering
    \includegraphics[width=0.75\linewidth]{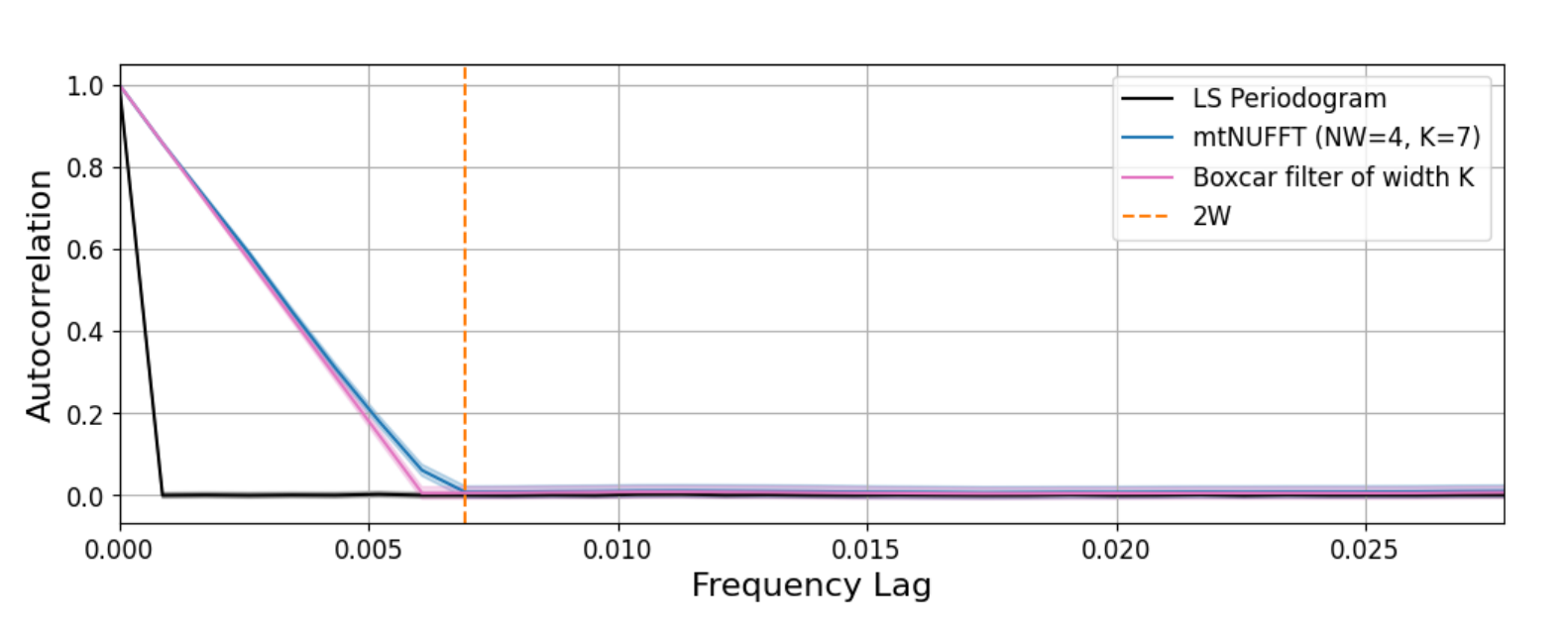}
    \caption{Auto-correlation as a function of frequency lag for LS, its filtered version using a moving-mean boxcar of width $K$, and \texttt{mtNUFFT} with $NW=4$ and $K=7$ parameters. We see that the \texttt{mtNUFFT} auto-correlation is similar to that of the moving-mean boxcar version of LS. Both of these power spectrum estimates are correlated within the $2W$ frequency range (marked by the orange dotted line), whereas the LS estimates not. All the periodograms have $\approx$0 correlations beyond this $2W$ range.}
    \label{fig:autocorrelation}
\end{figure*}

\texttt{mtNUFFT} power spectrum estimates separated by less than $2W$ in frequency are known to be correlated with each other. This correlation between adjacent estimates drops off to $\approx$0 beyond $\pm 2W$. We see such correlation trends in Figure 3, Page 344 of \cite{fitzgerald_2000} and Figure 4(a) of \cite{thomson_2014b}. To test this, we compute the auto-correlation of the \texttt{mtNUFFT} power spectrum estimate as a function of frequency lag for a toy-model time-series of random noise. We sample this time-series at the same times as Kepler-91. Then, we repeat this experiment 1000 times to get a distribution of the \texttt{mtNUFFT} auto-correlation for our toy-model. This auto-correlation is shown in Figure \ref{fig:autocorrelation}, along with that for the LS periodogram.  In this figure, we also plot the correlations introduced by a moving-mean boxcar filter of width $K$ bins applied to the LS periodogram, i.e., in the frequency domain. Here $K$ represents the number of tapers used to compute the \texttt{mtNUFFT} periodogram. We see that the \texttt{mtNUFFT} power spectrum estimates are correlated with each other within the frequency range of $2W$. Beyond this range, they are almost uncorrelated. The same is true for the filtered version of LS using the moving-mean boxcar. LS on the other hand is approximately uncorrelated across frequency bins.

\end{document}